\documentclass[11pt]{article}

\pdfoutput=1 
\usepackage{jheppub} 

\usepackage{graphicx}
\usepackage{amsmath, amssymb}

\usepackage{tikz}

\usepackage{verbatim}
\usepackage{caption}
\usepackage{subcaption}

\newcommand{\be}{\begin{eqnarray}}
\newcommand{\ee}{\end{eqnarray}}
\newcommand{\bea}{\begin{eqnarray}}
\newcommand{\eea}{\end{eqnarray}}

\newcommand{\bn}{\begin{enumerate}}
\newcommand{\en}{\end{enumerate}}

\title{$D$-type Conformal Matter and $SU/USp$ Quivers}

\author[a,d]{Hee-Cheol Kim,}
\author[b]{Shlomo S. Razamat,}
\author[a]{Cumrun Vafa,}
\author[c]{and Gabi Zafrir}

\affiliation[a]{Jefferson Physical Laboratory, Harvard University, Cambridge, MA 02138, USA}
\affiliation[b]{Department of Physics, Technion, Haifa, 32000, Israel}
\affiliation[c]{IPMU,  University of Tokyo,  Kashiwa, Chiba 277-8583, Japan}
\affiliation[d]{Department of Physics, POSTECH, Pohang 790-784, Korea}

\emailAdd{heecheol1@gmail.com}
\emailAdd{razamat@physics.technion.ac.il}
\emailAdd{vafa@physics.harvard.edu}
\emailAdd{gabi.zafrir@ipmu.jp}

\abstract{ We discuss the four dimensional models obtained by compactifying a single  M5 brane probing $D_{N}$ singularity (minimal D-type $(1,0)$ conformal matter in six dimensions) on a torus with flux for abelian subgroups of the $SO(4N)$  flavor symmetry. We derive the resulting quiver field theories in four dimensions by first compactifying on a circle  and relating the flux to duality domain walls in five dimensions. This leads to novel ${\cal N}=1$ dualities in 4 dimensions which arise from distinct five dimensional realizations of the circle compactifications of the D-type conformal matter.}

\begin{document} 
\maketitle
\flushbottom



\

\

\

\section{Introduction} 

The study of the dynamics of four dimensional supersymmetric quantum field theories has yielded examples of many interesting IR phenomena. These include, but are not limited to, symmetry enhancement, where the IR theory has a larger global symmetry than that expected from the UV description, and duality, where different UV descriptions flow to the same IR SCFT. Recently, with the advent of supersymmetric localization, we have increasingly more tools to conjecture and test examples of these phenomena via exact computation of various partition functions.

Having established such examples, a natural next step is to seek an organizing principle, from which all these phenomena arise naturally. One such a setup is to realize the $4d$ theory through the compactification of a $6d$ SCFT. As a simple example consider compactification of the $(2,0)$ theory on a torus leading to ${\mathcal N}=4$ super Yang-Mills in $4d$. As is well known this $4d$ theory is conformal and has a $1$ dimensional conformal manifold which in the compactification construction is realized as the complex structure of the torus. This space is spanned by a single complex variable $\tau$ defined in the upper half-plane. However, it is known that values of $\tau$'s differing by a modular transformation in fact define the same torus and thus the same compactification. This imply that $4d$ theories whose value of the coupling constant, $\tau$, differs by an $SL(2,Z)$ transformation should define the same SCFT. This is the well known S-duality of ${\mathcal N}=4$ super Yang-Mills, which is a dynamical non-perturbative phenomena of the theory. Yet, we see that it can be easily motivated from the geometrical properties of tori, by realizing it as the torus compactification of a $6d$ theory.  

This idea was later on extended to ${\mathcal N}=2$ theories by Gaiotto, by considering the compactification of the $(2,0)$ theory on more general Riemann surfaces \cite{Gai}. The class of theories constructed in this way is known as class S theories, and this method has been used to better understand the landscape and dynamics of ${\mathcal N}=2$ theories. For instance it can be used to motivate Argyres-Seiberg \cite{Argyres:2007cn} type dualities, and to construct various ${\mathcal N}=2$ SCFTs without a manifestly $\mathcal{N}=2$ preserving Lagrangian (the so called non-Lagrangian theories).    

A natural next step then is to try and apply this to ${\mathcal N}=1$ theories. One way is to consider again compactifications of the $(2,0)$ theory, but now only preserving ${\mathcal N}=1$ supersymmetry. This leads to the so called ${\mathcal N}=1$ class S theories that have been studied by various people, see for example \cite{Benini:2009mz, Bah:2012dg}. However, we can also consider starting from less supersymmetric theories, particularly, $6d$ $(1,0)$ SCFTs. These are far more numerous than their $(2,0)$ cousins, and many also posses interesting global symmetries that can be exploited in the compactification, and further should lead to $4d$ theories inheriting these symmetries.

This also ties neatly to another recent development in the study of $6d$ SCFTs, leading to many interesting results about these theories. There has been an intensive study aimed at classification of $(1,0)$ SCFTs from both F-theory compactifications and UV completion of gauge theories \cite{HMRV,ZHTV,Bhardwaj:2015xxa}, which led to a better understanding of the spectrum of $6d$ SCFTs. There also has been work on studying the t' Hooft anomalies of $(1,0)$ SCFTs \cite{OSTY}, and we now have tools to compute them.  Also studied is the compactification of $(1,0)$ SCFTs to $5d$, where they may have a low-energy description as a $5d$ gauge theory \cite{OSTY1,HKLTY,Zaf2,HKLY,OS,HKLTY1}. These recent developments set the stage for the study of the compactification of $(1,0)$ SCFTs to $4d$, and will be employed in this article for that purpose.

At this point in time, the study of the compactification of $(1,0)$ SCFTs to $4d$ has already been undertaken for specific choices of $(1,0)$ SCFTs and Riemann surfaces 
\cite{Gaiotto:2015usa,OSTY1,OSTY2,DelZotto:2015rca,Zaf2,OS,Razamat:2016dpl,Bah:2017gph,MOTZ,Kim:2017toz}. In this article we shall continue to expand the landscape of known compactifications of $(1,0)$ SCFTs to four dimensions. The specific case that we shall consider here is that of the minimal $(D_{N+3},D_{N+3})$ conformal matter compactified on a torus, or a sphere with two punctures, with fluxes in its global symmetry. 

The method employed to determine the field theories is to first reduce to $5d$ where the theory has an effective description as a $5d$ gauge theory. In fact it has at least three different effective descriptions\cite{HKLY}, and in this paper we will study two of those three with the remaining one to be discussed in a companion paper \cite{affpuyt}. The flux is then realized using a duality domain wall interpolating between two such descriptions. As the theory is compactified one direction is either a circle, in the case of the torus, or an interval, in the case of a two-punctered sphere, with boundary conditions at the two edges that play the role of the punctures. The resulting $4d$ theory can then be read off from the $5d$ bulk matter, compactified on intervals, and interacting through $4d$ fields living on the domain wall. This type of construction was successfully used to study \cite{Kim:2017toz} compactifications of the rank $1$ E-string, and here we generalize it to the case of the minimal $(D_{N+3},D_{N+3})$ conformal matter. The E string is the first case in this set of models, the $(D_4, D_4)$ minimal conformal matter, with the $SO(16)$ global symmetry enhancing to $E_8$.

Once a conjecture for the $4d$ theory was generated using the $5d$ picture, we can put that conjecture to the test by performing various consistency checks. Notably, the $6d$ reduction leads to various predictions that should be satisfied by the $4d$ theory. The most direct of which are the global symmetries and its 't Hooft anomalies, which can be computed by integrating the $6d$ anomaly polynomial on the Riemann surface \cite{Benini:2009mz}.   Moreover inequivalent realizations of the circle compactifications to $5d$ as well as moving the fluxes around lead to novel $4d$ duality predictions.  We performed these checks on many of these theories, and in any case we considered the theories we find are in agreement with these conditions.     

The structure of this article is as follows. We begin in section $2$ by presenting the $6d$ SCFT known as the minimal $(D_{N+3},D_{N+3})$ conformal matter. In particular we review some of its properties that are relevant for the later sections. We also use the $6d$ anomaly polynomial to predict the anomalies of the $4d$ theories resulting from the compactification of this $6d$ SCFT on a Riemann surface. Section $3$ is devoted to the five dimensional story. Here we consider first reducing the $6d$ SCFT to $5d$ and then employ the $5d$ low-energy gauge theory description to formulate a conjecture for the $4d$ theories. Section $4$ deals with the four dimensional story. Here we study the theories conjectured in the previous section, and compare the results against the $6d$ expectations.   Some aspects of the constructions are postponed to the appendices.   In particular, there is an alternative way to think of the minimal $(D_{N+3},D_{N+3})$ conformal matter in such a way that is naturally generalizable to the non-minimal as well as more general cases. This is part of a much larger story that will be considered in a different publication \cite{affpuyt}, but when applied to the case of the minimal $(D_{N+3},D_{N+3})$ conformal matter, it leads to other $4d$ field theories that should lead to dual descriptions of the same $4d$ SCFT.   We discuss this aspect in appendix B. 

\

\

\section{Six dimensions} \label{sec:6d}

We start by describing several known facts about the $6d$ SCFT called the $(D_{N+3},D_{N+3})$ minimal conformal matter. First we shall start with how this theory is constructed. There are various different ways to construct this SCFT in string theory. One way is as a ${\mathbb Z}_{2N-2}$ orbifold of the E-string theory\cite{MOTZ}. Alternatively it can be realized as the theory living on a single M5-brane probing a ${\mathbb C}^2/D_{N+3}$ singularity \cite{ZHTV}. In F-theory, it can be constructed by a $-1$ curve decorated with a $USp(2N-2)$ gauge group \cite{HMRV}. One interesting aspect of this theory is that it can be thought of as an orbifold of both the E-string theory and the theory living on a single M5-brane. Therefore, it is naturally related to both the orbifold generalizations of the $(2,0)$ theory and the E-string theory. In this article we shall mostly adopt the first viewpoint and think of it as a generalization of the E-string. However, the second viewpoint also exists and can be studied in a similar treatment, which we preform in a different publication \cite{affpuyt}. Nevertheless, we shall remark in appendix B on some interesting dualities upon comparing the two approaches.   

 This SCFT has a one-dimensional tensor branch along a generic point of which the SCFT reduces to a $USp(2N-2)$  gauge theory with $(2N+6)$ flavors. This leads to another useful description of the SCFT as the $6d$ UV completion of this gauge theory. We shall mostly employ this description to uncover various properties of this SCFT though we should state that most of them can also be seen from the other perspectives as well. This is important to bear in mind as properties of the $6d$ SCFT and its related IR gauge theory may differ \cite{OSTY2}. 

The first important property we shall need is the global symmetry. From the gauge theory we can see that there is an $SO(4N+12)$ global symmetry. Another useful property to keep in mind is that the theory has a moduli space, the Higgs branch, associated with giving vevs to the hypermultiplets. Generic points on this  initiates a flow that leads us from one value of $N$ to lower values. Specifically one can break the $USp(2N-2)$ group entirely in this manner. However the theory then is not empty as one still has the tensor multiplet. The string theory construction suggests that the resulting theory is the rank one E string theory. This is one way in which one can see that the $(D_{N+3},D_{N+3})$ minimal conformal matter can be thought of as a generalization of the rank one E-string, which from the field theory viewpoint is done by the addition of vectors and hypermultiplets. We also note that the naive expected symmetry $SO(16)$ in the E-string case is enhanced to $E_8$. We shall later discuss the mechanism leading to this enhancement from the gauge theory viewpoint.  

The $6d$ SCFT should then contain the conserved current of the $SO(4N+12)$ global symmetry. This is one important operator we observe already at the level of the gauge theory. There is one more interesting state, charged under the $SO(4N+12)$ global symmetry, that exists in the SCFT. The gauge theory contains non-perturbative excitations associated with instanton strings. These are massive at finite gauge coupling, but become massless at the origin of the tensor branch. This type of states is thought to play an important role in the UV completion. The gauge theory contains fermions, and these have zero modes in the instanton background which cause the instanton configuration to acquire flavor charges. From instanton counting one discovers that these instantons should then be in a chiral spinor of the $SO(4N+12)$ global symmetry. These lead to an additional BPS operator in the $6d$ SCFT that turns out to be a Higgs branch generator. Several aspects of the Higgs branch of this $6d$ SCFT, including the existence of this Higgs branch generator were discussed in \cite{HM}. We next review how this can be seen by compactifying the $6d$ SCFT to lower dimensions on circles. 

\

\subsection*{Circle compactification to lower dimensions} 

We start with the compactification of this theory to five dimensions on a circle. This has been analyzed by various people in different contexts \cite{OSTY1,HKLTY,Zaf2,HKLY,OS}, and here we collect some observations that will prove important later on. First we consider the problem of reduction to $5d$ on a circle with finite radius. We shall also allow turning on arbitrary holonomies for the $SO(4N+12)$ global symmetry, which correspond to additional mass parameters in the IR $5d$ theory. This problem was studied in \cite{HKLTY,HKLY}, where it was found that this theory reduces to some $5d$ gauge theory. The $5d$ gauge theory is not unique and in fact there are at least three different possible $5d$ gauge theories one can obtain depending on the holonomies one turns on.

The two descriptions which will prove most useful to us in this paper have only a single $5d$ gauge group. One description is a $5d$ $USp(2N)$ gauge theory with $2N+6$ fundamental  hypermultiplets while the other is a $5d$ $SU(N+1)_0$  gauge theory (subscript denotes the Chern-Simons level which is $0$) with  $2N+6$ flavors. These only have one gauge group so they only involve one large mass that is identified with the coupling constant. From the reduction viewpoint, this mass is identified with the radius, possibly tuned with an holonomy. There is another description involving a quiver gauge theory, $4F+SU(2)^{N}+4F$. This involves $N$ gauge couplings implying that we need at least $N-1$ holonomies to reach it. This also explains why the global symmetry is quite broken in it. We will not discuss this description here however this is the description which is the easiest to generalize to $ADE$ minimal conformal matter and we will address these general setups in a separate paper \cite{affpuyt}.

We next want to consider taking the zero radius limit with no holonomy. This limit was studied in \cite{Zaf2}, where it was found that the theory reduces to a $5d$ SCFT. It is convenient to allow mass deformations for this SCFT causing it to flow to an IR gauge theory. We can get at least three different IR gauge theories depending on the choice of mass deformations. These are related to the three different $5d$ descriptions of the $6d$ SCFT by integrating out a flavor\footnote{For example consider the case of the rank one E-string theory, where the three description degenerate. Compactifying on a finite circle, this $6d$ SCFT flows to an $SU(2)$ gauge theory with eight flavors assuming a suitable holonomy is turned on. It is known that when we take the zero radius limit the $6d$ SCFT flows to the $5d$ rank one $E_8$ SCFT\cite{GMS}. This $5d$ SCFT has a mass deformation where it flows to the $5d$ $SU(2)$ gauge theory with seven flavors, which is related to the former theory by integrating out a flavor.}. For our considerations it will suffice to concentrate on one of them, $USp(2N)$ gauge theory with $2N+5$ flavors. It is interesting to study the BPS spectrum, particularly the Higgs branch chiral ring, of this theory as this can teach us about the operator spectrum of the original $6d$ SCFT. In our case, a detailed analysis of the Higgs branch was preformed in \cite{FHMZ} primarily using the $3d$ mirror quiver one gets when compactifying the $5d$ SCFT to $3d$. We next review some aspects of this analysis while referring the reader interested in more in depth study of the Higgs branch to the reference.

 We can study the BPS spectrum of the $5d$ SCFT effectively using the $5d$ superconformal index \cite{Kim:2012gu}. One contribution we get comes from the hypermultiplets. Particularly the mesons are part of the $SO(4N+10)$ conserved current multiplet. Besides these there are no other independent invariants. Another important type of states are the instanton particles. These are non-perturbative excitations of the $5d$ gauge theory that become massless at the SCFT point. One important contributions of these operators is that they provide additional conserved currents that enhance the classically visible $U(1)_I \times SO(4N+10)$ global symmetry to $SO(4N+12)$ which is the symmetry of the $5d$ SCFT that is inherited from $6d$. These are expected to come from two-instanton contribution\cite{BZ}.

So we see that we get the $SO(4N+12)$ conserved current multiplet from the mesons, gaugino bilinear\footnote{This provides the scalar in the $U(1)_I$ conserved current multiplet.} and the two-instanton sector. This is identified with the reduction of the $6d$ conserved current multiplet, and naturally is in the adjoint of $SO(4N+12)$ and in the $\bold{3}$ of $SU(2)_R$. There is one more important contribution coming from the $1$-instanton sector. This provides a BPS operator in the spinor of $SO(4N+10)$. Together with the anti-instanton these form a chiral spinor of $SO(4N+12)$ that is in the $\bold{N+2}$ of $SU(2)_R$. This state is naturally identified as coming from the contribution of the instanton strings of the $6d$ SCFT.

\begin{figure}
\center
\includegraphics[width=0.7\textwidth]{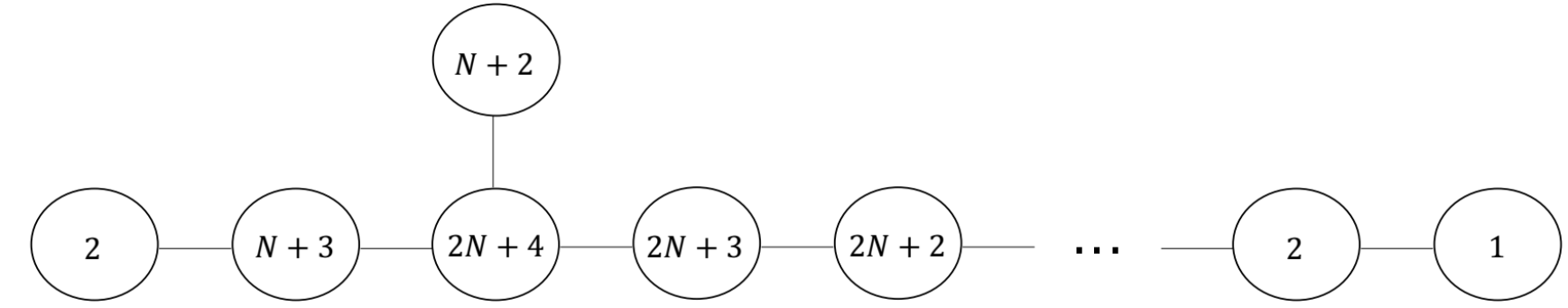} 
\caption{The $3d$ mirror dual of the SCFT one gets by compactification of the $6d$ SCFT $(D_{N+3},D_{N+3})$ conformal matter on $T^3$ without fluxes.}
\label{MirrorDual}
\end{figure}

We can further compactify to $4d$ where the $5d$ SCFT reduces to an A-type class S theory. From here it is also straightforward to reduce to $3d$ where we have a Lagrangian mirror dual \cite{BTX}. The specific dual for this case is shown in figure \ref{MirrorDual}. We first note that except the leftmost node, all other nodes are balanced. Thus from the results of \cite{GW}, we expect the symmetry on the Coulomb branch to enhance to $SO(4N+12)$ by monopole operators. These again match the $6d$ $SO(4N+12)$ conserved current multiplet. Next we turn to the unbalanced node. It too does contribute a monopole operator, but as it is unbalanced it won't be a conserved current. It is known that an unbalanced node would contribute a monopole operator in the $\bold{N_F-2N_C+3}$ of $SU(2)_R$\footnote{Here we refer to the R-symmetry that acts on the Coulomb branch.}, and in the representation of the global symmetry corresponding to the node in the Dynkin diagram it is connected to. For the case at hand this means we have a monopole operator in the $\bold{N+2}$ of $SU(2)_R$ and in the chiral spinor of $SO(4N+12)$.

So we see from $6d$, $5d$ and $3d$ perspectives that the basic operators charged under the $SO(4N+12)$ global symmetry are the conserved current, in the adjoint, and a Higgs branch generator in a chiral spinor. We do not appear to observe additional operators, particularly in $6d$, but also in lower dimensions. Importantly we do not observe operators in the vector and the other chiral spinor. This means that there is no contradiction with the $6d$ SCFT global symmetry group being more precisely $Spin(4N+12)/{\mathbb Z}_2$. Here we remind the reader that $Spin(4N+12)$ has a ${\mathbb Z}_2 \times {\mathbb Z}_2$ center where each element acts as $-1$ on the vector and one of the chiral spinors. The ${\mathbb Z}_2$ we mod by is the one not acting on the chiral spinor that appears in the SCFT. This will be important next when we discuss the reduction with fluxes.

\subsection*{$6d$ expectations from the reduction} 

In this section we shall discuss toroidal compactification of the $(D,D)$ minimal conformal matter to $4d$ with fluxes under its global symmetry. Particularly we shall consider the computation of the anomalies of the resulting $4d$ theory from those of the mother $6d$ theory. For that we require the anomaly polynomial of the $(D,D)$ conformal matter. This was computed in \cite{OSTY}. Alternatively it can be computed directly from the   $USp(2N-2)$   gauge theory description on the tensor branch. Either way it is found to be:

\bea
I_{DDMCM} & = & \frac{N(10 N + 3)}{24}C^2_2 (R) - \frac{N(2N+9)}{48} p_1 (T) C_2 (R) - \frac{N}{2} C_2 (R) C_2(SO(4N+12))_{V} \nonumber \\  & + & \frac{(N+2)}{24} p_1 (T) C_2(SO(4N+12))_{V} + \frac{(2N+1)}{24} C^2_2(SO(4N+12))_{V} \nonumber \\  & - & \frac{(N-1)}{6} C_4(SO(4N+12))_{V} + (29+(N-1)(2N+13))\frac{7p_1 (T) - 4p_2 (T)}{5760} \label{APDDCM}
\eea
We use the notation $C_2 (R)$ for the second Chern class in the fundamental representation of the $SU(2)_R$. We also employ the notation $C_n(G)_{\bold{R}}$ for the n-th Chern class of the global symmetry $G$, evaluated in the representation $\bold{R}$ (here $V$ stands for vector), and $p_1 (T), p_2 (T)$ for the first and second Pontryagin classes respectively.

Next we consider compactifying the theory on a torus with fluxes under $U(1)$ subgroups of $SO(4N+12)$. For simplicity we shall only consider the case of flux to a single $U(1)$ though the generalization to more $U(1)$'s is straightforward.
A basis for such choices is given by $2N+6$ distinct $U(1)$'s inside $SO(4N+12)$. These are just given by the Cartan subalgebra of $SO(4N+12)$. To each $U(1)$ we can associate a node in the Dynkin diagram of $SO(4N+12)$. Then for each node we get a different embedding of a $U(1)$ inside $SO(4N+12)$ where the commutant of the $U(1)$ in $SO(4N+12)$ is given by the Dynkin diagram one is left with after removing that node.

By examining the Dynkin diagram one sees that the possible embeddings preserve $U(1)\times SU(r) \times SO(4N+12-2r)$ for $r=1,2,...,2N+4$ with $r=2N+6$ being special. The special thing in the $r=2N+6$ case is that there are two $U(1)$'s associated with this case. This choice corresponds to the $U(1)$'s associated with the spinor nodes in the Dynkin diagram and so appear twice. What distinguishes the two choices is how the spinors decompose under $U(1)\times SU(2N+6)$. The group $SO(4N+12)$ has two inequivalent self-conjugate spinor representations, and under the embedding of $U(1)\times SU(2N+6) \subset SO(4N+12)$, one decomposes to all the even rank antisymmetric tensor representations of $SU(2N+6)$ while the other decomposes to all the odd rank ones. The two embeddings differ by which spinor decomposes to each choice. As we can exchange the two spinors by an outer-automorphism transformation on the generators of $SO(4N+12)$, the two embedding truly differ if the theory is not invariant under this transformation. Here we note that in the previous section it was demonstrated that the $(D,D)$ conformal matter $6d$ SCFT has a state in one of the chiral spinors of $SO(4N+12)$, but not the other. How this spinor decomposes under the global symmetry preserved by the flux differs between the two embeddings which in principle is distinguishable in the $4d$ theories. Therefore, we expect there to be two distinct flux choices, leading to distinct $4d$ theories, both preserving $U(1)\times SU(2N+6)$. As we shall see later, this leads to some rather surprising expectations in $4d$.

We next need to decompose the $SO(4N+12)$ characteristic classes into those of the commutant and the first Chern class of the $U(1)$. In this case we find that:

\bea
C_2(SO(4N+12))_{V} & = & - r C^2_1 (U(1)) + C_2(SO(4N+12-2r))_{V} + 2 C_2(SU(r))_{F} , \label{CCDecom} \\ \nonumber
C_4(SO(4N+12))_{V} & = & \frac{r(r-1)}{2} C^4_1 (U(1)) + C^2_2(SU(r))_{F} + 2 C_2(SU(r))_{F} C_2(SO(4N+12-2r))_{V} \\ \nonumber & & - r C^2_1 (U(1)) C_2(SO(4N+12-2r))_{V} + 2(3-r) C^2_1 (U(1)) C_2(SU(r))_{F} \\ \nonumber & & - 6 C_1 (U(1)) C_3(SU(r))_{F} + C_4(SO(4N+12-2r))_{V} + 2 C_4(SU(r))_{F}.  
\eea
Here $C_1 (U(1))$ is the first Chern class of the $U(1)$, normalized as in Appendix C. Using this we can next compute the anomalies of the resulting $4d$ theories.  

Before moving on to the anomaly calculation, we wish to introduce a flux basis. Fluxes can be associated with vectors on the root lattice so a basis of the root lattice can be used as a basis for fluxes. For the case of $SO(4N+12)$ this is simply given by a $2N+6$ vector built from the roots of $SO(4N+12)$, which are given by: $(\pm 1, \pm 1, 0 ,0 ,0 , ... ) + \text{permutations}$. In this basis a convenient choice to represent the fluxes we introduced is given by $( \overbrace{\pm z , \pm z , ... , \pm z}^{r} , 0 , 0 , ... , 0)$, where there are other choices related by Weyl transformations. The $U(1)$'s associated with the spinor nodes are both given by $r=2N+6$, but differ in whether the number of minus signs is even or odd. 

We expect that a flux is consistent if and only if it can be written as a sum of the roots of $SO(4N+12)$ with integer coefficients\footnote{Here when we refer to consistent flux we mean one consistent without introducing center fluxes.}. As a result, if $r$ is odd then $z$ must be even as one cannot build the $z$ odd vector using the $SO(4N+12)$ root vectors. This is related to the difference in the quantization condition between $r$ even and odd noted in Appendix C. It should be noted that the issue of flux quantization can be quite subtle due to the potential non-triviality of the global symmetry. Particularly, the `only if' part in the initial statement would be true if the group was $Spin(4N+12)$. However, if the group is $Spin(4N+12)/{\mathbb Z}_2$ then some apparently disallowed fluxes are possible as the states that made them inconsistent do not exist. For instant the flux associated with the spinor node whose associated spinor is in $Spin(4N+12)/{\mathbb Z}_2$ can have half-integer $z$. So, if we chose to associate it with the case of even number of minus signs, then the flux $(\frac{1}{2}, \frac{1}{2} , \frac{1}{2}, ... , \frac{1}{2})$ is consistent.      

\subsubsection*{Anomalies with flux}

Next we can consider compactifying the $6d$ theory on a Riemann surface $\Sigma$ with flux under the $U(1)$, that is $\int_{\Sigma} C_1 (U(1)) = -z$ where $z$ is an integer. For simplicity let us concentrate on the case where $\Sigma$ is a torus. As the torus is flat we do not need to twist to preserve SUSY. However, SUSY is still broken down to $\mathcal{N}=1$ in $4d$ by the flux. The $4d$ theory inherits a natural $U(1)_R$ R-symmetry from the Cartan of the $SU(2)_R$ though this in general is not the superconformal R-symmetry. Under the embedding of $U(1)_R \subset SU(2)_R$, the characteristic classes decompose as: $C_2 (R) = -C^2_1 (U(1)_R)$. 

Next we need to decompose $SO(4N+12)$ to the subgroup preserved by the flux, done in (\ref{CCDecom}). Finally we set: $C_1 (U(1)) = -z t + \epsilon C_1 (U(1)_R) + C_1 (U(1)_F)$. The first term is the flux on the Riemann surface, where we use $t$ for a unit flux two form on $\Sigma$, that is $\int_{\Sigma} t = 1$. The second term takes into account possible mixing of the $4d$ global $U(1)$ with the superconformal R-symmetry, where $\epsilon$ is a parameter to be determined via a-maximization. Finally the third term is the $4d$ curvature of the $U(1)$.   

Next we plug these decompositions into (\ref{APDDCM}) and integrate over the Riemann surface. This yields the $4d$ anomaly polynomial six form. From it we can read off the anomalies and find:

\be
Tr(U(1)_R) = Tr(U(1)^3_R) = Tr(U(1)_R U(1)^2_F) = 0, Tr(U(1)_F U(1)^2_R) = 2 N r z \label{AnomR}
\ee

\be
Tr(U(1)_F) = -2 r (N+2) z , Tr(U(1)^3_F) =-(3r + 2N - 2) r z,
\ee

\be
Tr(U(1)_R SO(4N+12)^2) = Tr(U(1)_R SU(r)^2) = 0 , Tr(SU(r)^3) = - 2 z (N-1), 
\ee

\be
Tr(U(1)_F SO(4N+12)^2) = - \frac{r z}{2} , Tr(U(1)_F SU(r)^2) = - \frac{(2N+r-2) z}{2} . \label{AnomG}
\ee
Here $U(1)_R$ refers to the $6d$ R-symmetry.
From this we can evaluate $a$ and determine $\epsilon$. We find that:

\be
\epsilon = sign(z) \frac{2\sqrt{5N+1}}{3\sqrt{2N+3r-2}} \label{Mix}
\ee
Using this we can evaluate $a$ and $c$:

\be
a= \frac{(5N+1)^{\frac{3}{2}} r |z|}{6 \sqrt{2N+3r-2}}, c = \frac{(11N+4) \sqrt{5N+1} r |z|}{12 \sqrt{2N+3r-2}},
\ee

As previously mentioned, for $r=2N+6$ there are two distinct choices of the embedding differing by how the spinor in the theory decomposes. However, the anomalies are indifferent to this distinction. Therefore, this suggests that there should be two distinct $4d$ theories with the same anomalies, but with slight differences in their matter spectrum.

\section{Five dimensions}\label{sec:5d}
As discussed in the previous section, circle compactifications of the $6d$ $(D,D)$ conformal matter lead to at least three different $5d$ gauge theories. In this section we study these $5d$ gauge theories when various fluxes are turned on under the compactifications.
The $6d$ fluxes naturally reduce to domain wall configurations in the $5d$ gauge theories \cite{Kim:2017toz}. We will call this type of domain walls as `flux domain walls'.
We shall construct Lagrangians of such domain walls coupled to the $5d$ systems that realize various fluxes along the global symmetry of the $6d$ $(D,D)$ conformal matter theory.

We split the $5d$ spacetime into two chambers and consider half-BPS interfaces interpolating between the two. Each chamber hosts a $5d$ gauge theory which is one of the three gauge theories from the $6d$ $(D,D)$ conformal matter possibly with different flavor holonomies. Two gauge theories and their boundary conditions will be connected by extra $4d$ degrees of freedom and superpotentials at the interface. When we properly choose the $4d$ fields in the domain wall as well as the $5d$ boundary conditions of the bulk $5d$ gauge theories and couple them through certain superpotentials, this domain wall can implement the flux domain wall of the $6d$ conformal matter theory.

\subsection*{$USp(2N)-SU(N+1)$ domain wall}
An interesting domain wall in the $5d$ reduction of the $6d$ $(D_{N+3},D_{N+3})$ conformal matter was studied in \cite{Gaiotto:2015una}. This domain wall glues together the $USp(2N)$ gauge theory on one side to the $SU(N+1)$ gauge theory on the other side. This domain wall is called `duality domain wall' as it interpolates between two dual gauge theories. This type of domain wall exists for any number of flavors, but we will focus on the cases with $N_f=2N+6$ fundamental flavors in both gauge theories. This domain wall is a higher rank $N>1$ generalization of the flux domain wall in the $5d$ E-string theory. When $N=1$ this domain wall implements a basic flavor flux in the $6d$ $(D_4,D_4)$ conformal matter theory (or $6d$ E-sting theory) as noted in \cite{Kim:2017toz}. Similarly, we conjecture that the rank $N$ duality domain wall with $N_f=2N+6$ is a flux domain wall of the $(D_{N+3},D_{N+3})$ conformal matter theory. More precisely, this domain wall corresponds to $1/4$ flux which breaks the $SO(4N+12)$ global symmetry down to $U(1)\times SU(2N+6)$ symmetry.

Let us briefly review the construction of the duality domain wall between two gauge theories of $USp(2N)$ and $SU(N+1)$ gauge groups in \cite{Gaiotto:2015una}. We will consider a 1/2 BPS interface located at $x^4=0$ along one of the spacial directions. We put the $USp(2N)$ gauge theory in the left chamber and the $SU(N+1)$ gauge theory in the right chamber. These gauge theories should satisfy 1/2 BPS boundary conditions at the interface. For the vector multiplets, we will choose Neumann boundary condition which sets the gauge field as
\be
	\partial_{4}A_{\mu}|_{x^4=0} = 0 \ \ (\mu=0,1,2,3) , \quad A_4|_{x^4=0} = 0 \ .
\ee

The boundary condition for the hypermultiplets is more involved. First, in the $USp(2N)$ gauge theory there are $2N+6$ fundamental hypermultiplets, $(X, Y^\dagger)^i\ (i=1,\cdots 2N+6)$ where $X^i$ and $Y^i$ are the fundamental chiral fields in the $i$-th hypermultiplet. For each $USp(2N)$ fundamental hypermultiplet, we have two choices of boundary conditions as either
\be\label{eq:boundarycondition-hyper}
	1)\;\;\, \ \partial_4X|_{x^4=0} = Y|_{x^4=0}=0 \quad \ {\rm or} \quad \ 2) \ X|_{x^4=0} =\partial_4 Y|_{x^4=0} = 0 \ .
\ee
We will denote this boundary condition by a sign vector $(\pm,\pm,\cdots,\pm)$ where $+$ or $-$ at the $i$-th entry stands for the first or the second boundary condition for the $i$-th hypermultiplet. So we have $2^{2N+6}$ different boundary condition choices for the $USp(2N)$ hypermultiplets. Each choice preserves a different $U(1)\times SU(2N+6)$ subgroup of the global symmetry.
On the other hand, the $SU(N+1)$ gauge theory has only two different choices of 1/2 BPS boundary conditions.
As we want to preserve $U(1)\times SU(2N+6)$ global symmetry, we should choose the same boundary condition for all the hypermultiplets. We thus have the boundary condition for the $SU(N+1)$ hypermultiplets as
\be\label{eq:boundarycondition-hyper-su}
 	1) \;\;\;\, \  \partial_4 X^i|_{x^4=0}=Y_i|_{x^4=0}=0 \quad \ {\rm or} \quad \ 2) \ \partial_4Y_i|_{x^4=0}=X^i|_{x^4=0}=0 \quad {\rm for \ all \ } i\ .
\ee
We will collectively call the chiral fields surviving at the interfaces as $M$ and $M'$ for the $USp(2N)$ and $SU(N+1)$ matters respectively.

The bulk $5d$ boundary conditions couple to the $4d$ boundary degrees of freedom at the interface such that the entire $5d$/$4d$ coupled system preserves $4$ real supersymmetries or $4d$ $\mathcal{N}=1$ supersymmetry. Since we give Neumann boundary conditions for the vector multiplets, the boundary $4d$ system has $USp(2N)\times SU(N+1)$ gauge symmetry.
We will introduce a bi-fundamental chiral multiplet $q$ of $USp(2N)$ and $SU(N+1)$ gauge groups and an anti-symmetric chiral multiplet $A$ of the $SU(N+1)$ gauge group.
We then couple these $4d$ fields to the $5d$ boundary conditions through the following 4d superpotentials:
\be
	\mathcal{W}|_{x^4=0} = M \cdot q \cdot M'+ q\cdot q \cdot A \ ,
\ee
where dot $\cdot$ denotes the contraction of the gauge and flavor indices in an appropriate manner.

This configuration provides a consistent $5d$/$4d$ coupled system. At the $4d$ interface, the $SU(N+1)$ gauge theory has in general cubic gauge anomaly of $N+3$ unit arising from the $2N+6$ chiral multiplets with the Neumann boundary condition. This cubic anomaly is canceled by the $4d$ boundary chiral fields $q$ and $A$ which contribute in total $-2N+(N-3)=-N-3$ to the cubic anomaly. As an anomaly free $5d/4d$ configuration, the above domain wall can naturally interpolate between the $USp(2N)$ gauge theory and the $SU(N+1)$ gauge theory coming from the $6d$ conformal matter theory on a circle.
There are three gauge anomaly free abelian symmetries.  One is the $6d$ $U(1)_R\subset SU(2)_R$ R-symmetry under which the bulk $5d$ fields $X$ and $Y$ transforms with charge $+1$. The $4d$ chiral fields $q$ and $A$ carry the $U(1)_R$ charge $0$ and $+2$ respectively. The gauge-$U(1)_R$ mixed anomaly coming from $q$ and $A$ is canceled by the contributions from the $5d$ vector multiplets with Neumann boundary conditions, while $X,Y$ do not contribute to this anomaly.
Another abelian symmetry is the flavor $U(1)_x$ symmetry acting on the fields $(M,M',q,A)$ with charges $(1,\frac{2}{N+1},-\frac{N+3}{N+1},\frac{2(N+3)}{N+1})$. This anomaly free $U(1)_x$ symmetry can in principle mix with the $U(1)_R$ R-symmetry.
The true R-symmetry of the low energy theory in the presence of the domain wall will be determined by a-maximization.
The last abelian symmetry is the $6d$ Kaluza-Klein (KK) symmetry. This symmetry remains unbroken even in the domain wall background by mixing with other abelian symmetries acting on the boundary fields. Under the $4d$ reduction which we will discuss below, this KK symmetry will become an extra $4d$ global symmetry when compactified on a finite size interval or will be broken when compactified on a torus.

\begin{figure}
\centering
\begin{subfigure}{.5\textwidth}
\centering
\includegraphics[width=0.55\textwidth]{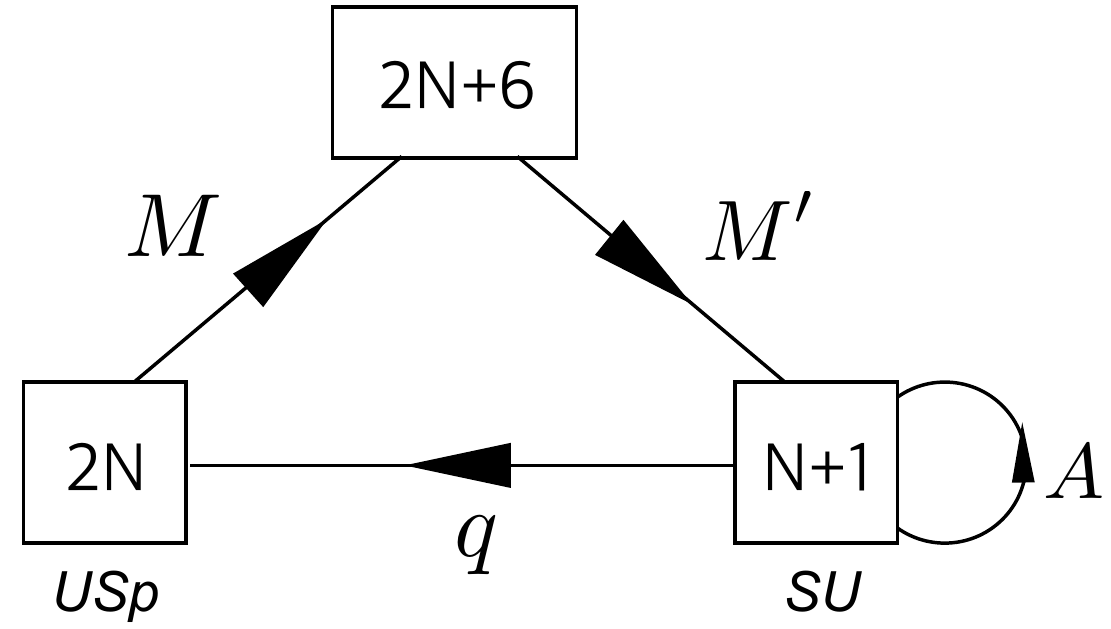}
\hspace{.5cm}
\caption{}
\label{fig:Sp-SU-wall}
\end{subfigure}%
\begin{subfigure}{.5\textwidth}
\centering
\includegraphics[width=0.45\textwidth]{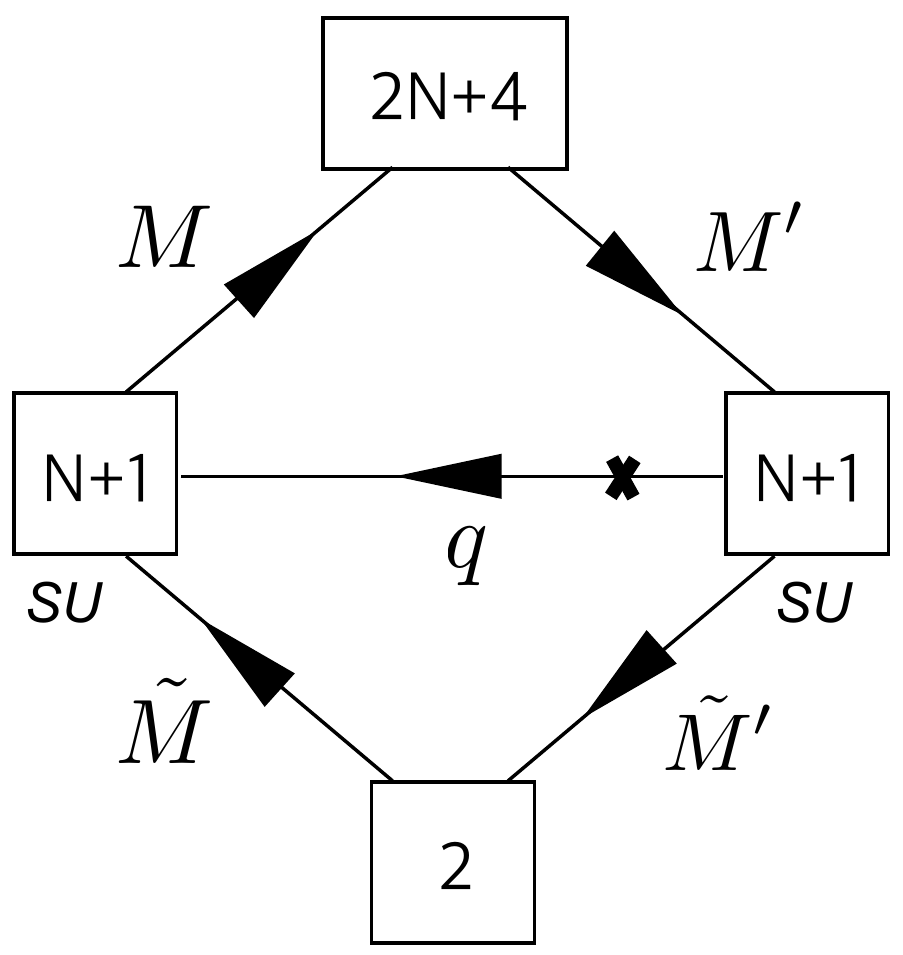} 
\caption{}
\label{fig:SU-SU-wall}
\end{subfigure}
\caption{Basic flux domain walls. Figure (a) is the $USp(2N)-SU(N\!+\!1)$ type domain wall with flux 1/4 preserving $U(1)\times SU(2N\!+\!6)$ symmetry. $USp(2N)\times SU(N\!+\!1)$ symmetry are gauged by the $5d$ bulk gauge groups. The chiral fields $M$ and $M'$ are from the hypermultiplets with Neumann boundary conditions on the two sides of the wall.
Figure (b) is the $SU(N\!+\!1)-SU(N\!+\!1)$ type domain wall with flux 1/2 preserving $U(1)\times SU(2)\times SU(4N+8)$ symmetry. $SU(N\!+\!1)\times SU(N\!+\!1)$ symmetry are gauged by the $5d$ bulk gauge groups. The chiral fields $M,\tilde{M}$ and $M',\tilde{M}'$ are from the $5d$ hypermultiplets with Neumann boundary condition. There is a gauge singlet chiral field denoted by `$X$' which couples to the baryonic operator of the bi-fundamental chiral field $q$.}
\end{figure}

So far we discussed the duality domain wall in \cite{Gaiotto:2015una} connecting the $USp(2N)$ and the $SU(N+1)$ gauge theories. The construction of the domain wall involves as described above the $5d$ boundary conditions, the extra $4d$ degrees of freedom, and the superpotential couplings.
We claim that this domain wall, which is drawn in Figure \ref{fig:Sp-SU-wall}, is the basic flux domain wall between the $USp(2N)$ and $SU(N+1)$ gauge theories. It introduces at the location of the domain wall the $1/4$ unit flux preserving $U(1)\times SU(2N+6)$ in the $6d$ $(D,D)$ conformal matter on a circle. In the orthogonal root basis, this flux corresponds to the flux along $(\frac{1}{4}, \frac{1}{4} , \frac{1}{4}, ... , \frac{1}{4})$. When there is only a single flux wall, all different choices of the boundary conditions for the hypermultiplets can be set to give the same flux by the Weyl symmetry of the $SO(4N+12)$ bulk global symmetry.
Note that this domain wall when $N=1$ reduces to the basic flux domain wall in the $5d$ E-string theory in  \cite{Kim:2017toz}.

We can carry out a simple but non-trivial check for this flux domain wall using a $4d$ reduction of this $5d$/$4d$ system. Basically, we will check that the $4d$ reduction of this system has the desired 't Hooft anomalies for being the $4d$ reduction of $6d$ $(D,D)$ conformal matter theory with fluxes. We first put this system in an interval $-L\le x^4 \le L$. In low energy below $\frac{1}{L}$, this system effectively reduces to a $4d$ theory. From our $5d$ Lagrangian description of the domain wall, we can deduce the $4d$ Lagrangian of the $6d$ ($D,D$) conformal matter theory put on a tube (or a two punctured sphere) with fluxes. The boundary conditions at $x^4=-L$ and $x^4=L$ define the punctures at both ends of the tube. We will choose the boundary conditions such that all chiral fields of $5d$ hypermultiplets with Neumann boundary condition at the domain wall (at $x^4=0$) also satisfy Neumann boundary condition at both ends.
The vector multiplets however are chosen to satisfy Dirichlet boundary condition at two ends of the tube. This truncates the $5d$ bulk gauge fields at the end of the each chamber, so the 5d gauge symmetries become non-dynamical global symmetries which we regard as global symmetries assigned to two punctures.
The $USp(2N)-SU(N\!+\!1)$ system with a basic flux domain wall then gives rise to a $4d$ Wess-Zumino type theory drawn in Figure \ref{fig:Sp-SU-wall} with $USp(2N)\times SU(N\!+\!1)$ global symmetries for the two punctures.

We claim the theory in Figure \ref{fig:Sp-SU-wall} is the $4d$ reduction of $6d$ $(D,D)$ conformal matter with flux $\frac{1}{4}$ on a tube with $USp(2N)\times SU(N\!+\!1)$ puncture symmetries. Let us check this proposal by comparing 't Hooft anomalies of this theory against direct computations from the $6d$ theory.
When we know a $5d$ Lagrangian description of a $6d$ theory, the $4d$ 't Hooft anomalies when compactified on a tube can be computed directly from the $6d$ anomaly polynomial together with anomaly inflow contributions coming from the $5d$ boundary conditions. See  \cite{Kim:2017toz} for detailed discussions. This computation does not rely on explicit Lagrangian descriptions of the $4d$ reduction. Integrating out the $6d$ anomaly polynomial (\ref{APDDCM}) on a tube with $1/4$ flux, we find the following 't Hooft anomalies:
\be\label{eq:anomaly-tube}
	&Tr(U(1)_R)=Tr(U(1)_R^3) = 0 \ , \ \ Tr(U(1)_RU(1)_x^2)= 0 \ , \ \  Tr(U(1)_xU(1)_R^2) = -N(N+3) \ , \nonumber \\
	& Tr(U(1)_x)=(N+2)(N+3) \ , \quad Tr(U(1)_x^3) = 4(N+2)(N+3) \ ,  \nonumber\\
	& Tr(U(1)_x\, SU(2N\!+\!6)^2) = \frac{N+1}{2} \ .
\ee
In addition, the 5d boundary conditions at both ends of the tube induce anomaly inflow contributions given by
\be\label{eq:anomaly-bc}
	&Tr(U(1)_R)=Tr(U(1)_R^3) = -\frac{3}{2}N(N\!+\!1) \ , \ \  Tr(U(1)_RU(1)_x^2)= Tr(U(1)_xU(1)_R^2) = 0 \ ,  \nonumber\\
	& Tr(U(1)_x)=2(N+1)(N+3) \ , \quad Tr(U(1)_x^3) = \frac{2(N+3)(N^3+2N^2+N+4)}{(N+1)^2} \ ,  \nonumber\\
	& Tr(U(1)_R\, USp(2N)^2) = -\frac{N+1}{2}\ , \quad Tr(U(1)_R\, SU(N\!+\!1)^2) = -\frac{N+1}{2} \ , \nonumber\\
	& Tr(U(1)_x\, USp(2N)^2) = \frac{N+3}{2} \ , \quad Tr(U(1)_x\, SU(N\!+\!1)^2) = \frac{N+3}{N+1} \ , \nonumber\\
	& Tr(U(1)_x\, SU(2N\!+\!6)^2) = \frac{N+1}{2} \ .
\ee
Combining (\ref{eq:anomaly-tube}) and (\ref{eq:anomaly-bc}), the result perfectly agree with 't Hooft anomalies of the $4d$ tube theory we proposed above. This provides a non-trivial evidence for our conjecture of the $4d$ tube theory and therefore for our flux domain wall conjecture given in Figure \ref{fig:Sp-SU-wall}. 

Domain walls for other general fluxes can be constructed by combining more than one basic domain wall with appropriate boundary conditions. We shall now explain how to connect two domain walls, which will be enough to construct general domain walls.
Suppose we locate two domain walls at $x^4=t_1$ and $x^4=t_2$ with $-L< t_1<t_2<L$. This splits the $5d$ spacetime into three chambers. Each chamber hosts either a $SU(N+1)$ gauge theory or a $USp(2N)$ gauge theory with $2N+6$ hypermultiplets. The $5d$ gauge theories in these three chambers are glued by two interfaces.

The boundary condition at each domain wall is the same as that of the single domain wall case: Neumann boundary conditions for vector multiplets and 1/2 BPS boundary conditions (\ref{eq:boundarycondition-hyper}) or (\ref{eq:boundarycondition-hyper-su}) for hypermultiplets. The $5d$ boundary condition couples the $4d$ chiral multiplets $q,A$ and $q',A'$ in the first and the second interfaces respectively with cubic superpotentials as we discussed. The theory in the second chamber lives in a finite interval with two boundaries. Thus, at low energy, the $5d$ theory in the second chamber reduces to a $4d$ gauge theory.
The hypermultiplets in the second chamber can have different boundary conditions at the two ends $t_1$ and $t_2$. When a hypermultiplet satisfies the same boundary conditions at both ends, it produces a $4d$ chiral multiplet coupled to the $4d$ gauge theory in the interval. On the other hand, a hypermultiplet obeying different boundary conditions at both ends becomes massive and will be integrated out at low energy. Integrating out this massive hypermultiplet at the end will induce a quartic superpotential between the chiral fields coupled to this hypermultiplet. This procedure defines our gluing rule between two or more flux domain walls.

The total flux of these two domain walls is determined by the boundary conditions of the hypermultiplets in the second chamber.
A domain wall turns on $+\frac{1}{4}$ flux along the $U(1)$'s rotating the chiral fields with Neumann boundary condition.
Thus, if a $5d$ hypermultiplet satisfies the same boundary condition at both ends, this combination of two domain walls introduces $\frac{1}{4}+\frac{1}{4}=\frac{1}{2}$ flux along the $U(1)$ acting on the hypermultiplet. However, if a hypermultiplet satisfies opposite boundary conditions at two ends, then the flux along the $U(1)$ direction is cancelled, $\frac{1}{4}-\frac{1}{4}=0$. The total flux is a sum of these $U(1)$ fluxes for the $2N+6$ hypermultiplets.

\begin{figure}
\centering
\begin{subfigure}{.5\textwidth}
\centering
\includegraphics[width=0.65\textwidth]{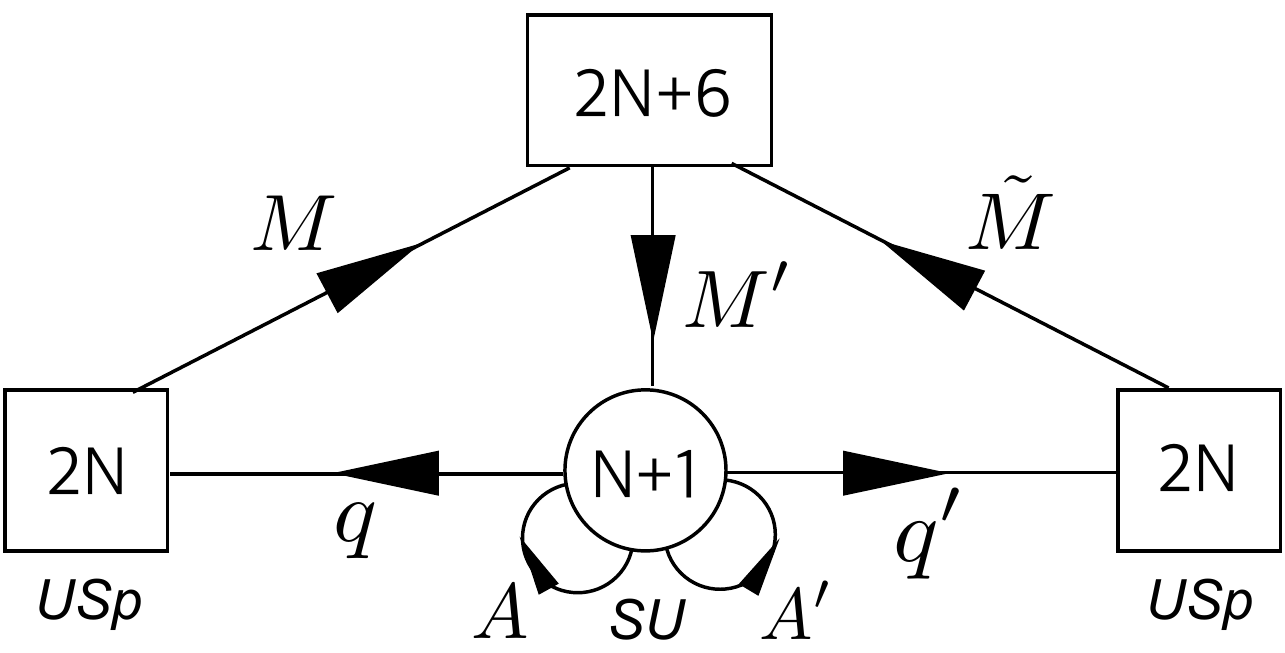}
\hspace{.5cm}
\caption{}
\label{fig:Sp-SU-Sp-wall}
\end{subfigure}%
\begin{subfigure}{.5\textwidth}
\centering
\includegraphics[width=0.75\textwidth]{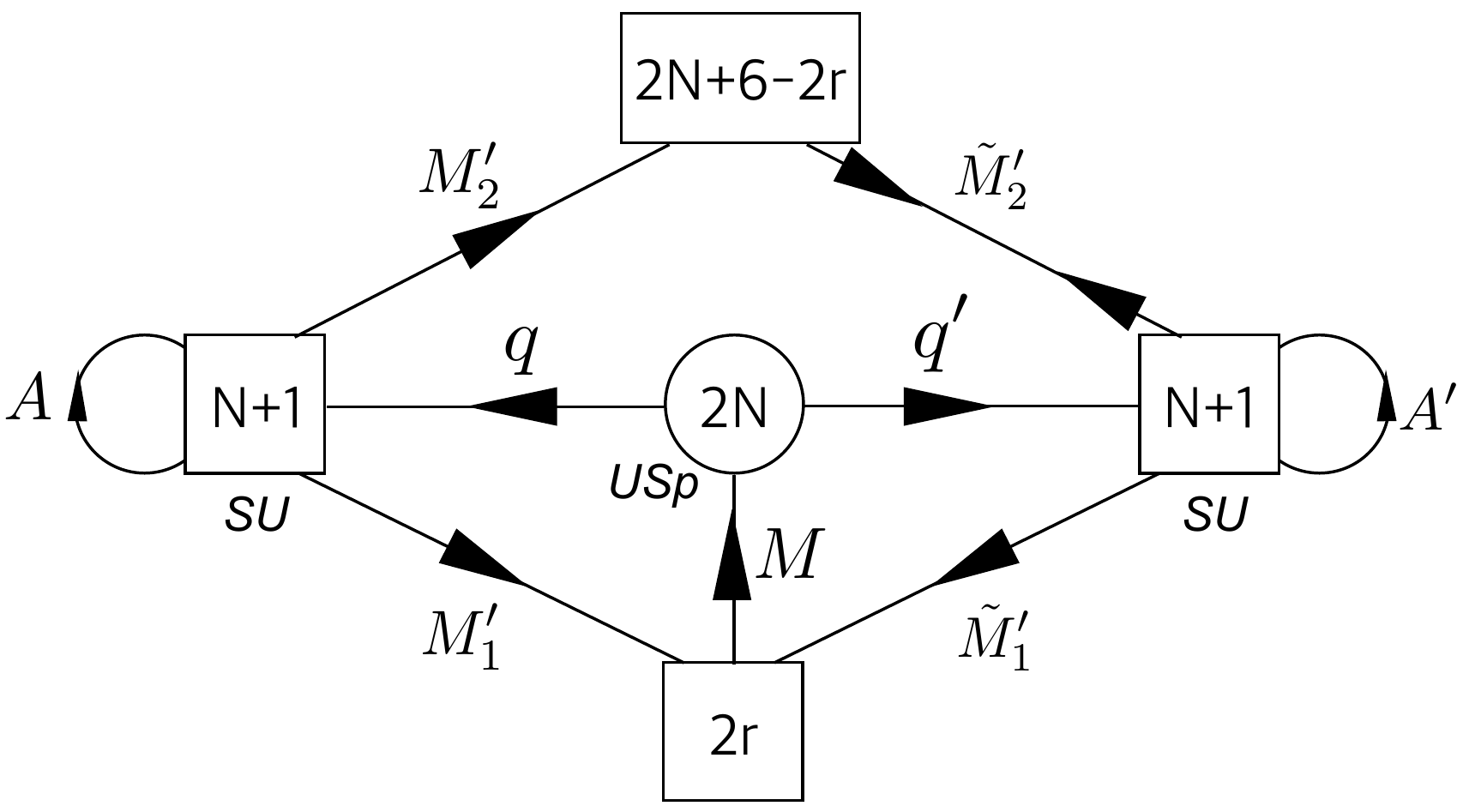} 
\caption{}
\label{fig:SU-Sp-SU-wall}
\end{subfigure}
\caption{Figure (a) is the domain wall connecting two $5d$ $USp(2N)$ gauge theories with flux $(\tfrac{1}{2},\tfrac{1}{2},\cdots ,\tfrac{1}{2})$.  Figure (b) is the domain wall connecting two $5d$ $SU(N+1)$ gauge theories with flux $(\frac{1}{2},\frac{1}{2},\cdots,\frac{1}{2} , 0 , 0 , ... , 0)$.}
\end{figure}

Let us begin with a domain wall configuration with $USp(2N)$ gauge theories in the first and the third chambers and $SU(N+1)$ gauge theory in the second chamber. In this case we have two different combinations with flux
\be\label{eq:two-domain-walls-flux}
	1) \  (\tfrac{1}{4},\tfrac{1}{4},\cdots ,\tfrac{1}{4}) +(\tfrac{1}{4},\tfrac{1}{4},\cdots ,\tfrac{1}{4}) \ , \quad 2) \ (\tfrac{1}{4},\tfrac{1}{4},\cdots ,\tfrac{1}{4}) +(-\tfrac{1}{4},-\tfrac{1}{4},\cdots ,-\tfrac{1}{4}) \ .
\ee
For the first configuration, we choose the boundary condition $(+,+,\cdots,+)$ for both $USp(2N)$ theories and $X_i=0$ (all $i$) for the $SU(N+1)$ theory at two interfaces. This boundary condition couples to the $4d$ boundary chiral fields $q,A$ and $q',A'$.
We will eventually obtain a flux domain wall configuration depicted in Figure \ref{fig:Sp-SU-Sp-wall}. This theory has two cubic superpotentials for each loop  in the quiver diagram and another two cubic superpotentials of the form $q^2A$ and $q'{}^2A'$ for the anti-symmetric $4d$ matters.
The $SU(N+1)$ gauge theory in the second chamber reduces to a $4d$ theory at low energy $E\ll (t_1-t_2)^{-1}$, and also the chiral halves of the $5d$ hypermultiplets satisfying Neumann boundary conditions at both boundaries become $4d$ chiral multiplets. In fact, we can regard this combination of two domain walls as a single domain wall coupled to the $4d$ $SU(N+1)$ gauge theory with chiral matter as shown in Figure \ref{fig:Sp-SU-Sp-wall}. This domain wall implements the flux $(\tfrac{1}{2},\tfrac{1}{2},\cdots ,\tfrac{1}{2})$ preserving $U(1)\times SU(2N+6)$ global symmetry of the $6d$ theory.

On the other hand, the second configuration in (\ref{eq:two-domain-walls-flux}) can be constructed with different boundary conditions. We choose  $(+,+,\cdots,+)$ for the first $USp(2N)$ theory at $x^4=t_1$ and  $(-,-,\cdots,-)$ for the second $USp(2N)$ theory at $x^4=t_2$. The hypermultiplets in the middle chamber satisfy $X=0$ at $x^4=t_1$ and $Y=0$ at $x^4=t_2$. Since the boundary conditions at the two ends are opposite, these hypermultiplets will be truncated at low energy while leaving a quartic superpotential of the form $M\,q q'\,\tilde{M}$. This gives rise to a trivial interface with zero flux. 

We now consider another type of two domain walls gluing $SU(N+1)$ gauge theories in the first and the third chamber and $USp(2N)$ gauge theory in the second chamber. Without loss of generality we can set the boundary conditions of the hypermultiplets in the $USp(2N)$ theory at $t_1$ as $(+,+,\cdots,+)$. So the first domain wall introduces a flux $(\frac{1}{4},\frac{1}{4},\cdots,\frac{1}{4})$. The boundary conditions at $t_2$ can be generically chosen as $(\overbrace{+,\cdots,+}^{2r} , -,\cdots,-)$. The $USp(2N)$ theory in the second chamber reduces to a $4d$ gauge theory with $2r$ fundamental chiral multiplets, say $M$, at low energy. These chiral multiplets come from the $5d$ hypermultiplets with Neumann boundary conditions at both ends. The number of chiral multiplets $M$ should be even due to the $Z_2$ anomaly of the $USp(2N)$ gauge group \cite{Witten:1982fp}.
The remaining $2N\!+6\!-\!2r$ hypermultiplets with opposite boundary conditions at the two ends are truncated in the IR. 
After all, this configuration reduces to the quiver diagram in Figure \ref{fig:SU-Sp-SU-wall}. Here, two $SU(N+1)$ symmetries are gauged by the $5d$ bulk gauge couplings. The domain wall has cubic superpotentials as $M_1'\,q\,M+\tilde{M}_1'\,q'\,M+q^2A+q'{}^2A'$ and a quartic superptential $M_2'\,qq'\,\tilde{M}'_2$. This domain wall corresponds to the $6d$ flux $(\overbrace{\tfrac{1}{2},\tfrac{1}{2},\cdots,\tfrac{1}{2}}^{2r} , 0 , 0 , ... , 0)$.

One interesting case is when $r=1$. In this case, we can simplify the quiver gauge theory on the domain wall using a known $4d$ duality. We can perform  Intriligator-Pouliot  duality \cite{hjk}  on  the $USp(2N)$ gauge theory in the second chamber. This leads to a rather simple Wess-Zumino type domain wall theory given in Figure \ref{fig:SU-SU-wall}. This is a flux domain wall with flux $(\frac{1}{2},\frac{1}{2},0,0,\cdots,0)$. We can treat this domain wall as another basic domain wall connecting two $5d$ $SU(N+1)$ gauge theories.

Generalization to more than two domain walls is straightforward.
We just consider a concatenation of different $5d$ gauge theories in several chambers connected by our flux domain walls discussed in this section. When we glue the two $5d$ theories we apply the above gluing rules, which include $5d$ boundary conditions, $4d$ chiral fields, and $4d$ superpotentials, at each interface such that the whole $5d$/$4d$ coupled system is consistent. The corresponding flux is automatically determined by the rules above. The total flux of the full $5d/4d$ system is the sum over fluxes of individual domain walls. We expect that each configuration realizes a $5d$ reduction of the $6d$ conformal matter theory with (generically different) flux. We note that the $6d$ flux does not depend on its location along $x^4$ direction. This suggests that there will be a large number of dualities between different domain wall configurations which give rise to the same total flux up to a Weyl transformation of the global symmetry. Under 4d reduction which we will study in the next section, the 5d domain wall dualities reduce to dualities between 4d $\mathcal{N}=1$ theories.

By putting the $5d$ systems with domain walls on an interval, we can obtain $4d$ Lagrangian theories of the $6d$ $(D,D)$ conformal matter theory compactified on a tube with general fluxes. For this we will give boundary conditions of the $5d$ theories at two ends of the interval appropriately and take the low energy limit. We choose Dirichlet boundary conditions for the vector multiplets. For the hypermultiplets, we give the 1/2 BPS boundary conditions such that chiral halves of the hypermultiplets survive in the first and the last chambers. This yields a $4d$ reduction of the $6d$ conformal matter with flux at low energy. The flux of the $4d$ theory is the same as the total flux of the domain walls. The 1/2 BPS boundary conditions define punctures at two ends of the interval. The puncture defined by this boundary condition hosts either $USp(2N)$ or $SU(N+1)$ global symmetry. 

We can also construct $4d$ theories of the $6d$ theory on a torus. We can glue the two ends of a tube by using the same gluing rules which we used to connect two or more domain walls. This will give rise to a $5d$ theory on a circle with flux domain walls.
At low energy this theory reduces to a $4d$ Lagrangian theory corresponding to the $6d$ $(D,D)$ conformal matter on a torus with flux.
In the next section, we will extensively test the $4d$ theories obtained from the $5d$ theories with flux domain walls compactified on an interval or a circle.

\

\section{Four dimensions}

\begin{figure}
\center
\includegraphics[width=0.7\textwidth]{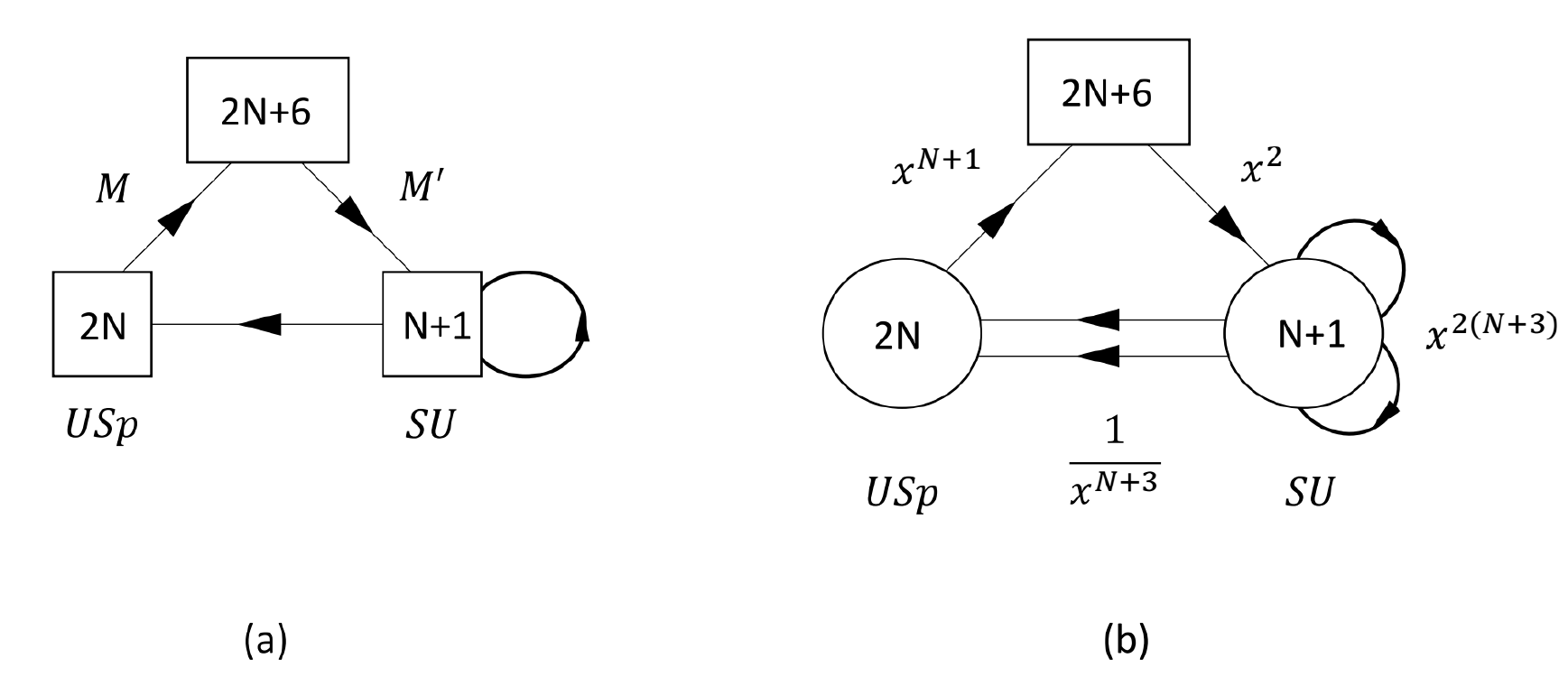} 
\caption{$4d$ theories corresponding to compactifications with flux preserving $SU(2N+6)$. Figure (a) shows the theory associated with a sphere with two punctures and flux $\frac{1}{4}$, while (b) shows the theory associated with a torus and flux $\frac{1}{2}$. The arrow from the $SU$ group to itself stands for an antisymmetric chiral field. There are cubic superpotentials for every triangle and for every antisymmetric chiral coupling it to two bifundamentals. The theory has an $U(1)_x \times SU(2N+6)$ global symmetry as well as a $U(1)_R$ symmetry. For the $U(1)_R$ it is convenient to use the $6d$ R-symmetry under which the bifundamentals have charge $0$, the antisymmetrics have charge $2$, and all the others have charge $1$. The charges under $U(1)_x$ are shown using fugacities.}
\label{BasicTube}
\end{figure}

We begin by studying the basic tube theory, which we associate with a sphere with two punctures and flux $\frac{1}{4}$ under the $U(1)$ breaking $SO(4N+12) \rightarrow U(1)\times SU(2N+6)$, where the spinor appearing in the $6d$ SCFT decomposes as the S spinor as defined in Appendix C. As argued in the previous section, the theory is just the collection of free fields with a superpotential shown in figure \ref{BasicTube} (a).

The tube has $SU(N+1)$, $USp(2N)$ and $SU(2N+6)$ global symmetries. It also has a $U(1)$ that remains unbroken when tubes are connected that we identify with the flux $U(1)$. The $SU(2N+6)$ is identified with the internal symmetry while the $SU(N+1)$ and $USp(2N)$ are associated with the two punctures. This implies that the two punctures are completely different, though have same rank, as the symmetries are not subgroups of one another. They arise as they are both $5d$ gauge groups that lift to the $6d$ $(D,D)$ conformal matter SCFT. This seems to suggests that there are at least two distinct variants of maximal punctures in this class of theories.  More generally for arbitrary 6d $(1,0)$ theory we expect there to be at least as many maximal punctures as inequivalent realizations of its 5d circle compactification.

There are chiral fields in the tube, denoted by $M$ and $M'$, that are charged only under the internal symmetries and puncture symmetries associated with one puncture. These play a role when we glue punctures as well as close them. We next consider gluing them.  Inspired by the structure in $5d$ we postulate that gluing the punctures is done by identifying and gauging their associated symmetries and introducing fundamental $2N+6$ chiral fields, $\Phi$, coupled via the superpotential $M \Phi + M' \Phi$. Essentially the punctures are glued using half the matter content of the corresponding $5d$ gauge theory.  Integrating out $\Phi$ would lead to identifying $M$ and $M'$ which is natural from the 5d viewpoint discussed in the previous section.  This happens when the boundary conditions on the two sides of the tube are the same.
This is the version of $\Phi$ gluing for this theory \cite{Razamat:2016dpl}, where we glue punctures of the same sign together. We expect there to be also a version of $S$ gluing \cite{Razamat:2016dpl}, we will discuss it in Appendix \ref{gk}.

We can now consider taking two tubes and connecting them together to form a torus. The resulting theory is shown in figure \ref{BasicTube} (b). We associate it with a compactification on a torus with flux $\frac{1}{2}$ preserving $SU(2N+6)$. This is the smallest theory we can get that corresponds to a torus\footnote{Naively, we cannot close the tube on itself since the two punctures are different. It should be noted that flux $\frac{1}{4}$ can be accommodated if we allow central fluxes in $SU(2N+6)$. It is interesting if the associated $4d$ theory can be found.}. We next perform a variety of checks on this theory.

We first start by noting some of its properties. Besides the $SU(2N+6)$, it also has a single non-anomalous $U(1)_x$, and a $U(1)_R$ symmetry, which can be identified with the Cartan of the $6d$ $SU(2)_R$. Under this $U(1)_R$ the bifundamentals have charge $0$, the antisymmetrics have charge $2$, and all other fields have charge $1$. This is identified with the $6d$ R-symmetry as $Tr(R) = Tr(R^3) = 0$. Therefore, the $4d$ global symmetry matches that expected from $6d$.

Next we can compare anomalies. We have already noted that the $6d$ R-symmetry anomalies match. We next wish to check the other anomalies. By comparing the anomalies with the expressions (\ref{AnomR})-(\ref{AnomG}) we find that these agree if we take $z=-\frac{1}{2}$ and $U(1)_x = (N+1) U(1)_F$. The first identification in particular suggests that the flux associated with it is indeed $\frac{1}{2}$. As explained in Appendix C, even though it is fractional, this flux is the minimal one possible without introducing central fluxes.   

We next consider the dynamics of this theory. We begin by looking for the superconformal R-symmetry of this theory. It should be written as $U(1)^{SC}_R = U(1)^{6d}_R + \alpha U(1)_x$. Performing a-maximization we find that:

\be
 \alpha=-\frac{\sqrt{5N+1}}{3(N+1)\sqrt{2(N+2)}}, \label{MixSim}
\ee 
as expected from (\ref{Mix}) and the mapping of the symmetries.

We can next analyze the flow to the IR and inquire what is the behavior of the theory at the deep IR. The first issue that arises is that computing the $\beta$ functions one finds that the $USp(2N)$ group is asymptotically free, but the $SU(N+1)$ group is IR free. Therefore, naively we may expect the $SU(N+1)$ group to become free at the deep IR, but one has to be careful since the $USp(2N)$ group do flow to strong coupling where the behavior of the $\beta$ function of the $SU(N+1)$ group may change.

To deal with this issue we consider first the same system, but without the $\mathcal{N} =1$ vector multiplet of the $SU(N+1)$ group, that is the same system in figure \ref{BasicTube} (b) but with the $SU(N+1)$ ungauged. The analysis we performed changes in two ways. First we do not have the contribution of the vector multiplet. This shifts a, but does not affect the a-maximization. 

The second change is that we are now no longer constrained by anomaly cancellation involving this $SU(N+1)$ group. Thus, in principle we may have new $U(1)$'s which can then mix with the R-symmetry changing the behavior of the flow. However, in this case we have no $U(1)$'s consistent with the superpotential and $USp(2N)$ anomaly cancellation so we conclude that the mixing for this theory is still given by (\ref{MixSim}).       

Now we note the following observations. First one can see that all gauge invariant operators are above the unitary bound. Thus, it is plausible that this theory flows to an SCFT in the IR. We  note that $Tr (U(1)^{SC}_R SU(N+1)^2) = -(N+1)$ so gauging the $SU(N+1)$ does not break $U(1)^{SC}_R$ implying that it is a conformal gauging. We expect that gauging the $SU(N+1)$ will give a new SCFT. We identify this SCFT with the IR theory that the quiver in figure \ref{BasicTube} (b) flows to.  

We next wish to explore the operator spectrum of this theory. Naturally it will be difficult to do so for any $N$. However we shall first present our expectations based on $6d$ reasoning and then compare them against the field theory. Finally we shall analyze a specific example, $N=2$, in detail using the superconformal index.

So first we ask what operators do we expect from $6d$. As discussed in section 2, the $6d$ SCFT contains two basic operators charged under the $6d$ global symmetry, one in the adjoint and the other in the spinor. Naively we expect that these should contribute operators in the $4d$ theory. We can easily read off their charges. First, as $U(1)^{6d}_R$ is just the Cartan of $SU(2)_R$, the operators coming from the adjoint of $SO(4N+12)$ should have the non conformal (six dimensional) R charge two in four dimensions, while those coming from the spinor should get R charge $N+1$. 

Their charges under the global symmetry could also be read off by decomposing the $SO(4N+12)$ into their $U(1)\times SU(2N+6)$ representations as given in the Appendix A. As previously mentioned when decomposing the $6d$ spinor state we have two different choices. These lead to a different operator spectrum in $4d$. We shall now argue by studying the spectrum of the $4d$ theory that the spinor here decomposes like the S spinor decomposition as stated in Appendix C\footnote{The two choices also differ by flux quantization, where only this choice allows half-integer flux without breaking the global symmetry.}.

But first let us consider the states in the adjoint of $SO(4N+12)$. These should give as operators in the adjoint of $SU(2N+6)$, as well as ones in the antisymmetric. The adjoint $SU(2N+6)$ operators just come from the triangle, that is the invariant made from a $USp$ flavor, an $SU$ flavor and a bifundamental. These are in the singlet and adjoint of $SU(2N+6)$ and are marginal operators in the $4d$ theory. 

We can also have an invariant made from two flavors of the $USp$ group. This is a marginal operator under the $6d$ R-symmetry, but not under the superconformal one as it is charged under $U(1)_x$. We can get another operator, marginal under the $6d$ R-symmetry, but with opposite charges, from the invariant made from two $SU(N+1)$ flavors and two bifundamentals. These two operators form the``off-diagonal" parts in the decomposition of the $SU(2N+6)$ adjoint.

 More invariants that can be built are baryons made just from $SU(N+1)$ flavors and antisymmetrics. Particularly we can consider an invariant made from $k$ antisymmetrics and $N+1-2k$ flavors. One can see that all of these have $6d$ R-charge $N+1$, $U(1)_x$ charge $2(N+1)(k+1)$, and in the rank $N+5+2k$ antisymmetric representation of $SU(2N+6)$. These precisely form the representations appearing in the decomposition of the spinor with positive $U(1)_x$ charge. We can also use the anti-fundamental of $SU(N+1)$, constructed from a $USp$ flavor and a bifundamental, and the anti-antisymmetric, constructed from two bifundamentals, and construct similar invariants. These give all the representations appearing in the decomposition of the spinor with negative $U(1)_x$ charge. The middle rank $N+3$ antisymmetric representation appears to be absent.

So far we have identified states that can be linked to a $6d$ operator. However that identification has been crude. We next want to consider one example and study the superconformal index. We shall see that the observations made before appear also in the index. For our test case we take $N=2$ which is the simplest case after the E-string. For the purpose of index calculations we shall use the R-symmetry $U(1)^{6d}_R - \frac{1}{10} U(1)_x$. Since $\frac{\sqrt{11}}{9\sqrt{8}} - \frac{1}{10} \approx 0.03$, this R-symmetry is close to the superconformal R-symmetry. We then find:

\bea
  \mathcal{I} & = & 1 + x^6 (p q)^{\frac{7}{10}} \chi[\overline{\bold{45}}] + 2 x^{12} (p q)^{\frac{9}{10}} \chi[\bold{10}] + \frac{1}{x^{12}} (p q)^{\frac{11}{10}} \chi[\overline{\bold{10}}] + x^6 (p q)^{\frac{7}{10}} (p + q) \chi[\overline{\bold{45}}] \\ \nonumber & + & x^6 (p q)^{\frac{6}{5}} \chi[\bold{120}] - \frac{1}{x^6} (p q)^{\frac{13}{10}} \chi[\bold{45}] + 2 x^{12} (p q)^{\frac{9}{10}} (p + q) \chi[\bold{10}] + x^6 (p q)^{\frac{7}{5}} (x^{12}(\chi[\bold{210}]+\chi[\bold{825}]) \\ \nonumber  & - & \frac{1}{x^{18}} \chi[\bold{8}]) + \frac{1}{x^{30}} (p q)^{\frac{3}{2}} + ... \label{IndexUSpSU}
\eea   

We can see the two operators $x^6 (p q)^{\frac{7}{10}} \chi[\overline{\bold{45}}]$ and $\frac{1}{x^6} (p q)^{\frac{13}{10}} \chi[\bold{45}]$. These are exactly the off-diagonal terms that appear in the decomposition of the adjoint. One can see that these contribute at order $p q$ under the $6d$ R-charge. We also note that their number is given as expected from \cite{BRZ} (see \cite{Kim:2017toz} Appendix E ). There are no terms at order $p q$ which again agrees with the expectations of \cite{BRZ}.

We also have the operators $x^{12} (p q)^{\frac{9}{10}} \chi[\bold{10}]$ and $x^6 (p q)^{\frac{6}{5}} \chi[\bold{120}]$ which we can identify as coming from the $6d$ spinor state. These are precisely the lowest order operators that appear in the decomposition of the $SO(20)$ spinor to $SU(10)$. One can also see that under the $6d$ R-charge, they contribute at order $(p q)^{\frac{3}{2}}$ as expected. We do note that we do not see the $\bold{252}$, which is expected at order $p q$. Incidentally, the number of these operators, including the $\bold{252}$, is given also by the formula of \cite{BRZ}.

\subsection*{Generalized tubes}

In this section we discuss tubes corresponding to flux under the $U(1)$ breaking $SO(4N+12) \rightarrow U(1)\times SU(r) \times SO(4N+12-2r)$ for $r$ even. The relevant tubes are shown in figure \ref{Tubesa} (a). We associate with these tubes flux $z=-\frac{1}{2}$, using the normalization conventions in Appendix C. The tubes have an $SU(r) \times SU(2N+6-r)$ non-abelian global symmetry, and two non-anomalous $U(1)$'s that remain after closing the tubes. We identify these with the $U(1)\times SU(r) \times SO(4N+12-2r)$ global symmetry expected from $6d$.

\begin{figure}
\center
\includegraphics[width=0.8\textwidth]{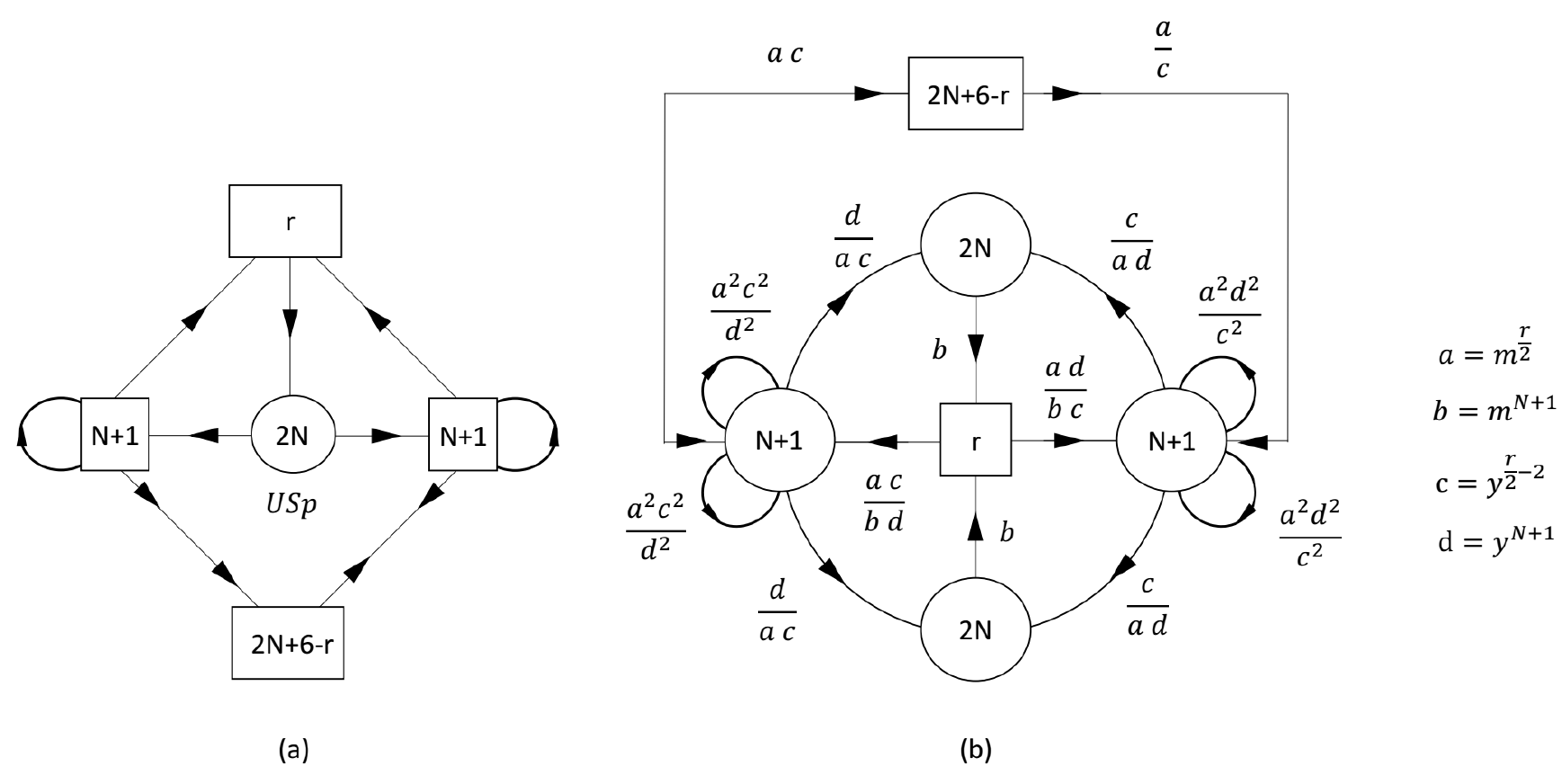} 
\caption{$4d$ theories corresponding to compactifications with flux preserving $SU(r) \times       U(1)\times  SO(4N+12-2r)$. All groups are $SU$ except the ones with $2N$ that are  symplectic,  $USp$. Here $r$ is even so that the $USp(2N)$ gauge group is not anomalous. For bifundamentals between $SU$ groups we adopted a double arrow notation indicating whether the field is in the fundamental or antifundamenal of each $SU$ group. Like before the arrow from the $SU$ group to itself stands for an antisymmetric chiral field. (a) The theory corresponding to a tube with flux $z=-\frac{1}{2}$. There are cubic superpotentials for every triangle and for every antisymmetric chiral coupling it to two bifundamentals. Additionally there is a quartic superpotential for the lower 'triangle'. (b) Connecting two tubes leads to this theory.}
\label{Tubesa}
\end{figure}

The tubes also have two $SU(N+1)$ global symmetries that we identify with the punctures. These tubes, thus, have two punctures of the same types. The bifundamentals connecting these puncture symmetries to the internal symmetries are identified with the fields $M$ and $M'$ that play a role in the gluing.  

We can consider taking two tubes and connecting them to form a torus with flux $z=-1$. The resulting theory is shown in figure \ref{Tubesa} (b). We associate this theory to a $U(1)\times SU(r) \times SO(4N+12-2r)$ preserving compactification on a torus with flux $z=-1$. The visible symmetries in the Lagrangian are $SU(r) \times SU(2N+6-r)\times U(1)_m \times U(1)_y$. We also have an R-symmetry, identified with the Cartan of the $6d$ $SU(2)_R$, with the fields charged as in the basic tube. The anomalies of this theory matches the ones computed from $6d$ with $z=-1$ if we identity $U(1)= (N+1) U(1)_m$ and $SO(4N+12-2r)\rightarrow U(1)_y \times SU(2N+6-r)$ such that $\chi[\bold{4N+12-2r}]_{SO(4N+12-2r)} \rightarrow y^{\pm(N+1)}\chi[\bold{2N+6-r}]_{SU(2N+6-r)} + \frac{1}{y^{\pm(N+1)}}\chi[\overline{\bold{2N+6-r}}]_{SU(2N+6-r)}$.

We next analyze some dynamical aspects of this theory, starting with the superconformal R-symmetry. It is not difficult to see that only $U(1)_m$ can mix with the $6d$ R-symmetry. Thus we set $U(1)^{SC}_R = U(1)^{6d}_R + \alpha U(1)_m$, where we find:

\be
  \alpha=-\frac{2\sqrt{5N+1}}{3 (N+1) \sqrt{2N+3r-2}}, \label{aGenT}
\ee 
as expected from (\ref{Mix}).

Like in the previous case we have the problem that the $SU(N+1)$ groups are IR free. However the $USp(2N)$ groups are asymptotically free in the  range of $r$. It is therefore possible that the flow of the $SU(N+1)$ groups is changed along the flow of the $USp(2N)$ groups. We can test this in a similar way as in the previous case. Here we further have the complication that there are quartic superpotentials that are irrelevant in the UV. We shall first ignore this issue and then return to it later. 

We consider the theory where the $SU(N+1)$ groups are global symmetries and inquire where such a theory flows to. One can show that relaxing the anomaly cancellation condition of the two $SU(N+1)$ groups, we have an additional $U(1)$ rotating the two $SU(N+1)\times SU(2N+6-r)$ bifundamentals with opposite charges. This $U(1)$ does not mix with the R-symmetry so the theory still has mixing given by (\ref{aGenT}). Particularly compared to that R-symmetry the $SU(N+1)$ groups have zero $\beta$ function and so gauging them does not initiates a flow.

\begin{figure}
\center
\includegraphics[width=0.38\textwidth]{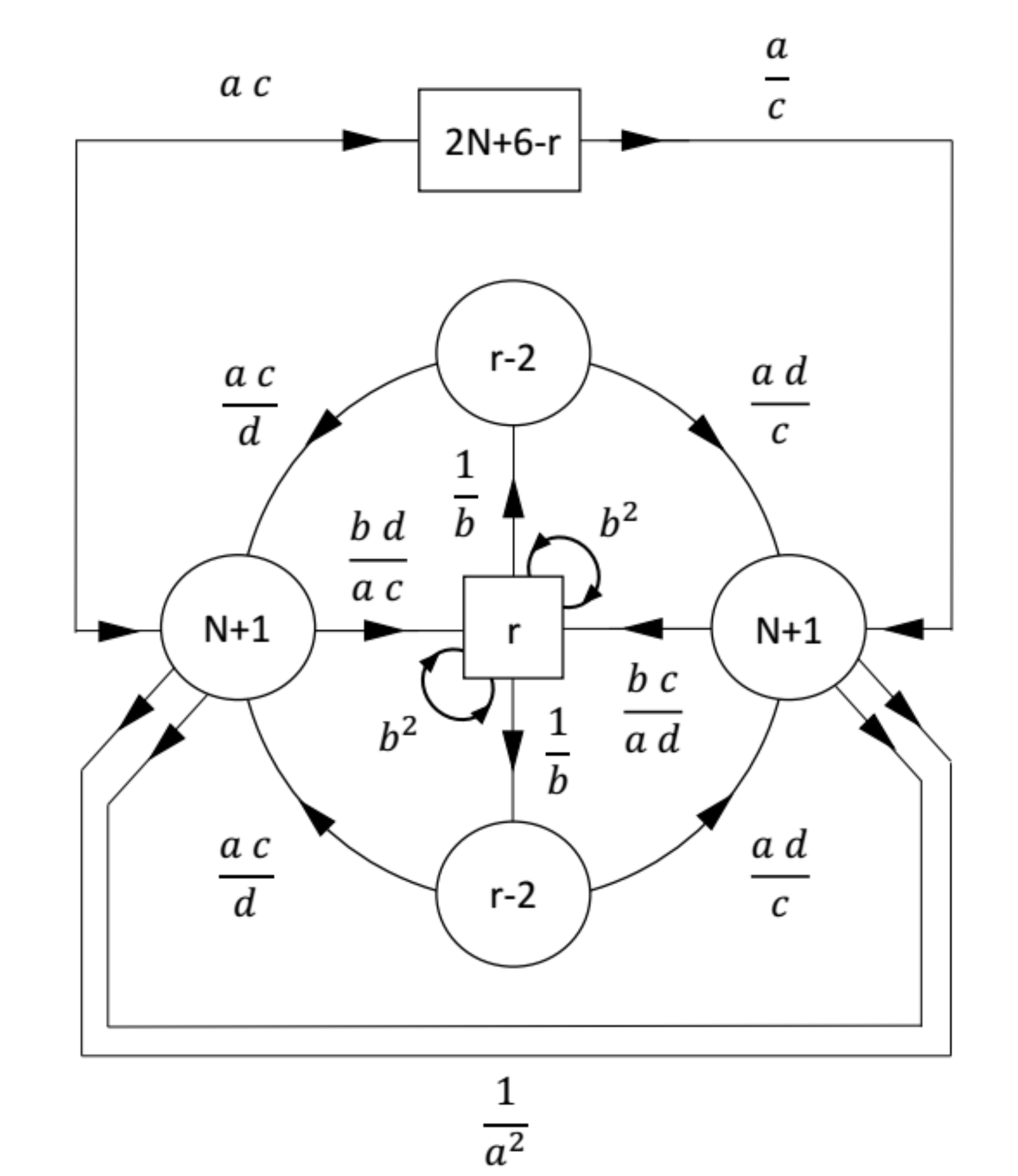} 
\caption{The theory we get after performing Intriligator-Pouliot duality on the two $USp$ groups in the theory in figure \ref{Tubesa} (b). Here the groups with $r-2$ are of type $USp$ and the various fugacities are related as stated in figure \ref{Tubesa}.}
\label{AfterDul}
\end{figure}

This leaves the question of the nature of the theory we flow to. Examining the operators dimension we find that for $r \geq N+1$ all gauge invariant operators appear to be above the unitary bound so an SCFT is plausible. However, when $r < N+1$ the $USp(2N)$ mesons go below the unitary bound. This can be attributed to the fact that the conformal window for a $USp(2N)$ gauge theory with $2N_f$ fundamental chiral fields ends when $N_f < \frac{3(N+1)}{2}$. We can then perform Intriligator-Pouliot duality \cite{hjk} to get to a description with IR free $USp(r-2)$ groups shown in figure \ref{AfterDul}. Recall that the theory has a superpotential coupling the $SU(N+1)$ antisymmeric and $SU(r)$ bifundamental to the $USp(2N)$ mesons. After the duality where the mesons get promoted to basic fields, these become mass terms, and many of these fields get integrated out. After the dust settles we find that the $SU(N+1)$ groups sees $2N+r+2$ effective flavors. Thus, for $r \geq N+1$, the $SU(N+1)$ are IR free and the $USp(r-2)$ group asymptotically free exactly like the dual description. However when $r < N+1$ the groups reverse roles in this description: the $USp(r-2)$ groups become IR free while the $SU(N+1)$ groups become asymptotically free. This means that gauging the $SU(N+1)$ groups is a relevant deformation in this range and will initiate a flow that may change the behavior of the $USp$ groups.  

We can proceed to analyze this theory in a somewhat similar manner. We first start by ungauging the $USp$ groups. In this case we do not get any new symmetries so the theory flows as before. As noted when $r < N+1$ some operators hit the unitary bound. In this description these are the $SU(r)$ antisymmetric chiral fields that flip some of the $USp(r-2)$ mesons. It is thus likely that these decouple and become free fields in the IR. We can proceed and try to analyze the theory while treating these fields as free. The result of the a maximization will now be different and in general is quite ugly. To simplify we can check some special cases. For instance in the first non-trivial case, $r=N$, we explicitly evaluated the R-symmetry, under the assumption that only the $SU(r)$ antisymmetric chiral fields become free, and we have not found operators violating the unitary bound. Therefore, at least for one case it is plausible that the theory flows to an SCFT plus free fields. In general we expect that as $r$ decreases the dimension of operators decreases and that other operators may hit the unitary bound. We have thus also checked the $r=4$ case, where we again do not find any unitary bound violating operators. 

\begin{figure}
\center
\includegraphics[width=0.28\textwidth]{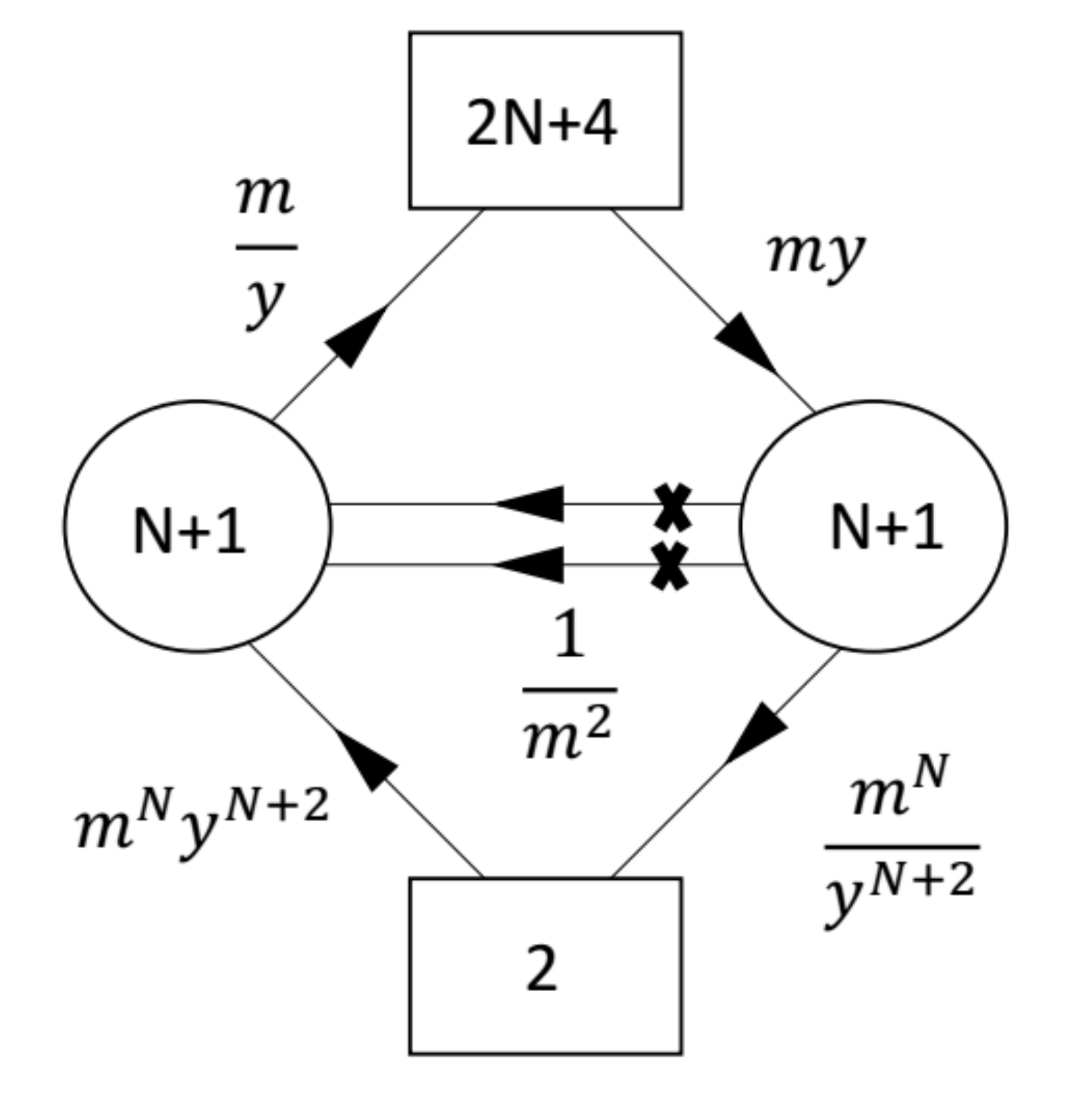} 
\caption{The theory we get after performing Intriligator-Pouliot duality on the two $USp$ groups in the theory in figure \ref{Tubesa} (b) when $r=2$. Here there is a cubic superpotential along the upper triangle and an order $N+2$ along the lower triangle. Additionally there are two singlet fields that flip the baryons made from the bifundamentals, which is represented by the 'X' drawn on the bifundamentals lines.}
\label{ThrRisTwo}
\end{figure}

The $r=2$ case is somewhat special as in this case we do not have $USp$ gauge groups in the dual description, and the dual theory simplifies to the theory in figure \ref{ThrRisTwo}. Now there are no IR free groups, but there is an irrelevant superpotential of the form $F_L B^{N} F_R$ associated with the lower triangle, as well as the irrelevant flipping superpotential. The later appears to stay irrelevant along the flow as the two flipping fields go below the unitary bound suggesting that they actually decouple in the IR. The former, however, appears to be marginal as turning it off does not lead to additional symmetries that can mix with the R-symmetry. We do gain an extra $SU(2)$ symmetry that should appear as an accidental symmetry at some point on the conformal manifold. Since there are two bifundamentals, and so more than one such kind of superpotential, we expect that some of these marginal operators should be exactly marginal.  

We can also consider analyzing the operator spectrum assuming only the flipping fields decouple. Again we do not find any other operator going below the unitary bound. So to conclude the discussion on the dynamics of these theories, we seem to find no contradiction with them flowing to an SCFT when $r > N+1$ and to an SCFT plus $r(r-1)$ free fields when $r \leq N+1$.

Finally we return to the issue of the quartic superpotentials. We can attempt to address both these and the IR free gauge couplings by turning off these superpotentials in addition to ungauging the $SU(N+1)$ gauge groups. In this case we do get an additional $U(1)$ that mixes with the R-symmetry so that the the $SU(N+1) \times SU(2N+6-r)$ bifundamentals have free R-charge, as these become decoupled free fields once both the $SU(N+1)$ gauge couplings and superpotentials are turned off. We can again perform a maximization, while regarding the $SU(N+1) \times SU(2N+6-r)$ bifundamentals as free fields. The results are again quite messy so we shall not write them down. However, one can show that with respect to that R-symmetry either the superpotential is relevant and the gauging is irrelevant or vice verse, with the only exception being $r=N+1$ where they are both marginal. It seems that when $r>N+1$ the gauging is relevant while the superpotential is irrelevant, and vice verse for $r<N+1$. Since turning any one of them breaks the additional $U(1)$ that mixes with the R-symmetry, as long as one of them is relevant the theory should flow to the interacting SCFT with the mixing as in (\ref{aGenT}). The $r<N+1$ is somewhat complicated by the fact that the some mesons of the $USp(2N)$ groups become free. In that case we have seen that the theory is easier to analyze from the dual frame where the $SU(N+1)$ groups are asymptotically free while the $USp(2N)$ groups are IR free. In that frame the quartic superpotential become cubic, and so we do not expect any modification from the results of the previous analysis. So to conclude, we see no indication that the quartic superpotentials decouple in a way that will modify the previous conclusion.  

Next we want to analyze some aspects of the spectrum. Particularly we will look at the various BPS states in the theory and try to find states expected to match those in $6d$. Naturally, this will be somewhat crude and we shall later try to study some cases more explicitly using the superconformal index. Nevertheless we shall see that there are operators one can naturally identify with those expected from $6d$ reasoning.

We expect to find $4d$ operators coming from both the $6d$ conserved current and the spinor state. Here the $6d$ global symmetry is $U(1)\times SU(r) \times SO(4N+12-2r)$ so we expect the index to form characters of $SO(4N+12-2r)$. Let's start with the conserved current multiplet. It should contribute operators with $6d$ R-charge $2$ and with charges as worked out in Appendix A. We can indeed identify these in the $4d$ theory. First we have the $SU(r)\times SU(2N+6-r)$ gauge invariant bifundamentals we can build from the $SU(r)\times USp(2N)$, $USp(2N)\times SU(N+1)$ and $SU(N+1)\times SU(2N+6-r)$ bifundamentals. This indeed has $6d$ R-charge $2$, $U(1)_m$ charge $N+1$, in the fundamental of $SU(r)$ and in the $y^{N+1}\chi[\bold{2N+6-r}]_{SU(2N+6-r)}$ and $\frac{1}{y^{N+1}}\chi[\overline{\bold{2N+6-r}}]_{SU(2N+6-r)}$. Looking at the mapping expected from anomalies, these match the operator expected from $(F_{SU(r)},V_{SO(4N+12-2r)})^1$. We also have the conjugate state from the $SU(r)\times SU(N+1)$ and $SU(N+1)\times SU(2N+6-r)$ bifundamentals connected through two $USp(2N)\times SU(N+1)$ bifundamentals.

The states in the antisymmetric of $SU(r)$, coming from the $SO(4N+12)$ conserved current, can be identified with the $USp(2N)$ mesons associated with the $SU(r)\times USp(2N)$ bifundamental. The conjugate is given from the $USp(2N)$  `meson' generated from the $SU(r)\times SU(N+1)$ and $SU(N+1)\times USp(2N)$ bifundamentals. This completes the states that we see from the  $6d$ conserved current.

We can also see some of the states associated with the spinor. For instance we can consider the baryons, from the right $SU(N+1)$, group made from $a$ $SU(r)\times SU(N+1)$ bifundamentals, $b$ $SU(2N+6-r)\times SU(N+1)$ bifundamentals and $c$ antisymmetrcs, where $a+b+2c=N+1$. All of these have $6d$ R-charge $N+1$, $U(1)_m$ charge $(N+1)(\frac{r}{2}-a)$, $U(1)_y$ charge $(N+1)(N+3-\frac{r}{2}-b)$ and are in the rank $r-a$ antiymmetric of $SU(r)$ and rank $2N+6-r-b$ antiymmetric of $SU(2N+6-r)$. Here $a$ and $b$ run from $0$ to $N+1$ or $r$ for $a$ and $2N+6-r$ for $b$ depending on which one is smaller. Also due to the constraint $a+b+2c=N+1$ the even or oddness of $a$ and $b$ are correlated. These states exactly match some of the states expected from the spinor. We also have the baryons from the left $SU(N+1)$ group, and the baryons made from the anti-antisymmetic, made from two $USp(2N)\times SU(N+1)$ bifundamentals, and antifundamentals we get from the $USp(2N)$ fundamentals and $USp(2N)\times SU(N+1)$ bifundamentals. The latter is forced to give the conjugate representations due to the chiral ring relations enforced by the superpotentials.

Finally we shall examine the superconformal index for some selected cases. We start with the case of $N=2$, $r=8$. From (\ref{aGenT}) we see that $U(1)^{SC}_R = U(1)^{6d}_R -\frac{\sqrt{22}}{9 \sqrt{13}} U(1)_m$. For the purpose of index calculation we shall employ the R-symmetry $U(1)^{6d}_R - \frac{1}{9} U(1)_m$, which is reasonably close to the superconformal one since $\frac{\sqrt{22}}{9 \sqrt{13}} - \frac{1}{9} \approx 0.03$. Evaluating the superconformal index we find:

\bea
  \mathcal{I} & = & 1 + 2 m^6 (p q)^{\frac{2}{3}} \chi[\bold{1},\bold{1},\overline{\bold{28}}] + (p q)^{\frac{5}{6}} (4m^{12} \chi[\bold{2},\bold{1},\bold{1}] + m^3 \chi[\bold{2},\bold{2},\overline{\bold{8}}]) + 3 m^9\chi[\bold{1},\bold{2},\bold{8}] p q + ... \nonumber \\ \label{IndexUSpSUNEXT}
\eea   
where we use $\chi[SU(2),SU(2)_y,SU(8)]$ and $\bold{2}_{SU(2)_y} = y^3 + \frac{1}{y^3}$. As can be seen from (\ref{IndexUSpSUNEXT}) the index can indeed be written in characters of the global symmetry expected from $6d$, $SU(2)\times SU(2)_y\times SU(8)\times U(1)_m$. Furthermore, all terms appearing in (\ref{IndexUSpSUNEXT}) have a natural $6d$ origin as expected from the reasonings of \cite{BRZ}  (see also \cite{Kim:2017toz} Appendix E) and the branching rules in Appendix C. Specifically, the terms $m^6 (p q)^{\frac{2}{3}} \chi[\bold{1},\bold{1},\overline{\bold{28}}]$ and $m^3 \chi[\bold{2},\bold{2},\overline{\bold{8}}]$ are the ones expected from the adjoint state, while the others are the ones expected from the spinor. We also note that the conserved current contributions exactly cancels against that of the marginal operators again in accordance with the $6d$ expectations.

We next consider the case of $N=r=2$. This case is easiest to approach from the dual description in figure \ref{ThrRisTwo}. From (\ref{aGenT}) we see that $U(1)^{SC}_R = U(1)^{6d}_R -\frac{\sqrt{11}}{9 \sqrt{2}} U(1)_m$. However, using this the two flipping fields have R-charges below the unitary bound so these must decouple. Preforming the a maximization again, under the assumption that these are free fields, we now find $U(1)^{SC}_R = U(1)^{6d}_R -\frac{5}{18} U(1)_m$. With this R-symmertry all gauge invariant operators have R-charges above the unitary bound. We next proceed to evaluate the superconformal index. For that, it is convenient to work with the R-symmetry $U(1)^{6d}_R -\frac{1}{4} U(1)_m$, instead of the superconformal one. Note that since $\frac{5}{18} - \frac{1}{4} \approx 0.03$, both R-symmetries are quite close. Also since the two singlets decouple in the IR, we shall ignore them in this calculation. Evaluating the superconformal index we find:

\bea
  \mathcal{I} & = & 1 + m^3 (p q)^{\frac{5}{8}} \chi[\bold{2},\bold{16}]	+ \frac{4}{m^6}(p q)^{\frac{3}{4}} + (\chi[\bold{3},\bold{1}]-1) p q + m^3 (p q)^{\frac{5}{8}} (p+q) \chi[\bold{2},\bold{16}] + m^3 (p q)^{\frac{9}{8}} \chi[\bold{1},\bold{128'}] \nonumber \\ & + & \frac{2}{m^6}(p q)^{\frac{3}{4}} (p+q) + m^6 (p q)^{\frac{5}{4}} (\chi[\bold{3},\bold{135}] + \chi[\bold{1},\bold{120}] - \chi[\bold{3},\bold{1}]) + \frac{1}{m^3} (p q)^{\frac{11}{8}} \chi[\bold{2},\bold{16}] +  \frac{10}{m^{12}}(p q)^{\frac{3}{2}} ... \nonumber \\ ,\label{IndexUSpSUNNEXT}
\eea
where here we write the index in characters of the global symmetry expected from $6d$, $SU(2)\times SO(16)\times U(1)_m$. We have ordered the character has $\chi[SU(2),SO(16)]$ and we have: $\bold{16}_{SO(16)} = y^3\bold{8}_{SU(8)} + \frac{1}{y^3}\overline{\bold{8}}_{SU(8)}$, $\bold{128'}_{SO(16)} = \frac{1}{y^9}\bold{8}_{SU(8)} + \frac{1}{y^3}\bold{56}_{SU(8)} + y^3 \overline{\bold{56}}_{SU(8)} +y^9 \overline{\bold{8}}_{SU(8)}$. 

Some of the operators appearing above have a $6d$ interpretation. Particularly the first term is the one expected from the $6d$ conserved current multiplet. This multiplet is also expected to give two singlets contributing as $m^6 (p q)^{\frac{1}{4}}$, which are exactly the two flipping fields that we ignored in this calculation. We also have the term $m^3 (p q)^{\frac{9}{8}} \chi[\bold{1},\bold{128'}]$ which matches part of the contribution expected from the spinor. The $U(1)_m$ independent contribution from the spinor is not observed, similarly to as in the previous cases. The remaining terms are either product of lower operators or ones that have no immediate $6d$ origin. Here, unlike the previous cases, the conserved currents and marginal operators do not cancel exactly. Particularly we have a marginal operator in the $\bold{3}$ of the $SU(2)$ that is not expected from $6d$. As a result the structure of the conformal manifold deviates from the $6d$ expectations. This operator comes from the superpotential term involving two bifundamentals and one $SU(3)\times SU(2)$ flavor from each of the $SU(3)$ gauge groups, where the $SU(2)$ contraction is done symmetrically. As we have two bifundamentals there are three different choices for this superpotentials. Two are canceled against the conserved currents of the two $SU(2)$ that would be there were the superpotentials turned off, leaving only the single contribution seen in the index. What we want to stress in this analysis is that this extra contribution is related to the existence of the two bifundamentals which is a property of the low flux and so is not generic and we expect the deviation with the $6d$ expectation to vanish once the flux increases. 

Finally we wish to analyze a case where $r=N+1$. Since $r$ must be even, the simplest new case is $N=3$, $r=4$. This case is easiest to analyze from the dual picture in figure \ref{AfterDul}, where the theory is manifestly free. We can evaluate the superconformal index for this theory. We note that here there are $12$ fields, flipping the mesons of the global $SU(4)$, that are decoupled. For simplicity we shall ignore them in the calculation. We find:

\bea
  \mathcal{I} & = & 1 + (p q)^{\frac{2}{3}} (\frac{2}{m^8}\chi[\bold{6},\bold{1}] + m^4 \chi[\bold{4},\bold{16}])	+ ... \label{IndexUSpSUDual}.
\eea

Here we have written the index in characters of the global symmetry expected from $6d$, $SU(4)\times SO(16)\times U(1)_m$. We have ordered the character has $\chi[SU(4),SO(16)]$ and we have: $\bold{16}_{SO(16)} = y^4\bold{8}_{SU(8)} + \frac{1}{y^4}\overline{\bold{8}}_{SU(8)}$. We have performed the analysis up to order $p q$ whose contribution is found to be vanishing.

\begin{figure}
\center
\includegraphics[width=0.38\textwidth]{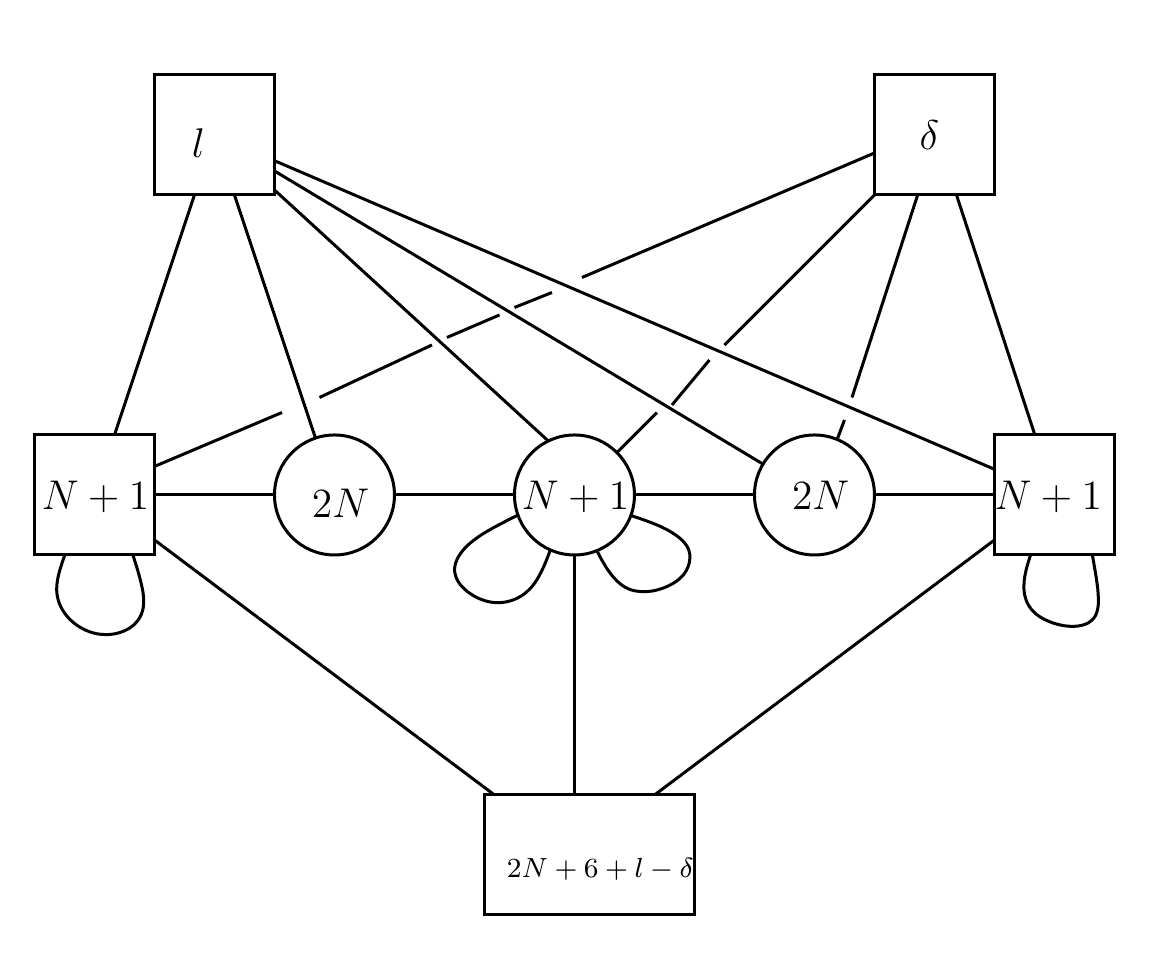} 
\caption{Theory obtained by gluing together two tubes with different flux. One is with $r=l$ and the other one with $l+\delta$. The lines will follow the obvious orientation pattern from 
Figure 5.}
\end{figure}

Of the two states appearing in the index, the last one can be identified as coming from the $6d$ conserved currents. The other state is eliminated from the chiral ring by the singlets, which provide the second contribution coming from the $6d$ conserved currents.  

We can consider gluing tubes with different flux together to obtain tubes with fluxes in more than single $U(1)$. This amounts to splitting the internal symmetries in the appropriate manner. When glued together we can compute anomalies of the models and compare them with six dimensional compactifications on a torus. We do find agreement though the details become rather cumbersome. We just write down the tube in Figure 8 which one gets by gluing tubes with two different fluxes. If one considers gluing two such tubes together the superconformal R symmetry is obtained by maximizing,

\be
&&a(s, h)=\frac{3}{32}(-3l s^3(3l+2N-2)+  
\nonumber\\
&&18 l\delta  h s^2+6\delta h^3(3\delta +2N-2)-9 \delta ls h^2+4l s (5N+1)-8\delta h(5N+1)\,.
\ee

\

\

\

\newpage

\noindent {{\bf Acknowledgmenets}}  
We would like to thank Patrick Jefferson for useful discussions.  We also like to thank SCGP summer workshop 2017 for hospitality during part of this work.
  The research of HK and CV is supported in part by NSF grant PHY-1067976. GZ is supported in part by  World Premier International Research Center Initiative (WPI), MEXT, Japan.  The research of SSR was  supported by Israel Science Foundation under grant no. 1696/15 and by I-CORE  Program of the Planning and Budgeting Committee.

\appendix

\section{S gluing}\label{gk}

Let us comment here on gluing punctures of opposite sign. The punctures break the $SO(4N+12)$ symmetry to $SU(2N+6)\times U(1)$. One can define the color of the puncture as the choice of the embedding of $SU(2N+6)$ in $SO(4N+12)$. The punctures come with operators $M$ charged under puncture symmetry and $SU(2N+6)\times U(1)$. Punctures of
opposite sign have operators in conjugate representations of these symmetries. We can glue punctures of different sign. The procedure is called S gluing, see for example \cite{Razamat:2016dpl} for definition in related set ups. To do so we introduce gauge fields for the puncture symmetry and turn on superpotential $M  M'$ coupling the operators coming from the glued punctures.

\begin{figure}[htbp]
\center
\includegraphics[width=0.48\textwidth]{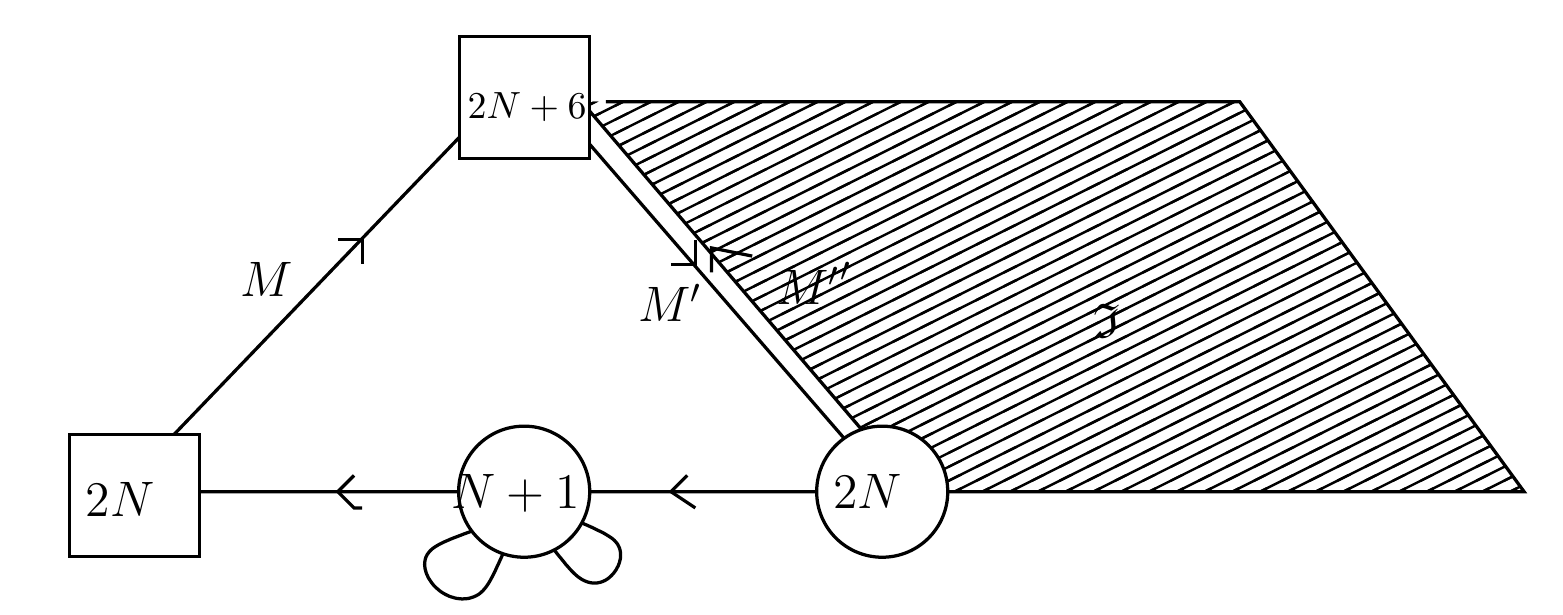} \includegraphics[width=0.42\textwidth]{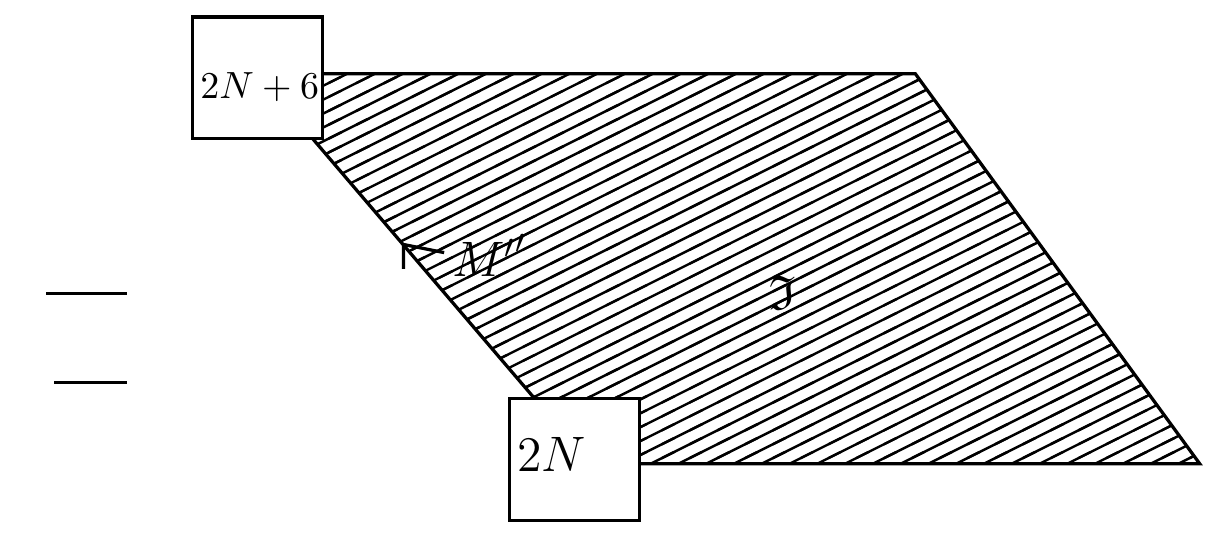} 
\caption{S gluing with $SU$ groups.}
\end{figure}

We can define tubes with punctures of opposite signs by conjugating all representations. It is natural to associate to these tubes opposite flux. One can then perform a consistency check by gluing two such tubes to a general model ${\frak T}$. The theory then should be the same as the model ${\frak T}$ because the value of flux does not change and the surfaces are the same after and before the addition of the tubes. 
We can verify that indeed the above occurs. The reason is that when two tubes of opposite flux are combined the superpotentials and matter content are consistent with triggering a vacuum expectation value which Higgses the gauge group. 

There are two cases we can consider. One is when we combine the tube models of opposite flux with $SU(N+1)$ gauging and other with $USp(2N)$ gauging. In both cases the mesonic operators (in $USp$ case they are in $(N+1)\times \overline{(N+1)}$ representation of two $SU(N+1)$ groups, and in  case of $SU$ they are in $2N\times 2N$ of two $USp$ groups) obtain vacuum expectation values, Higgsing the group following which the theory flows back to the model ${
\frak
 T
}$. In the latter case the vacuum expectation is triggered by quantum effects following from analysis of the $USp$ gauge theory \cite{hjk}, and in the former  the superpotential  with the fields in antisymmetric and the bifundamentals is needed to trigger the flow. For example in case of $N=2$ the $SU(3)$ gauging has five fundamental flavors, as the antisymmetric is anti-fundamental here. Without the superpotential there is no vacuum expectation value generated, but with it the claim is that it will be. One piece of evidence in favor of this is that such a vacuum expectation value, inthe presence of the superpotential, is not forbidden by symmetries. It would be interesting to understand whether it is actually turned on 
by instanton effects or not, and we conjecture for consistency of our picture that it is.
 
 \begin{figure}[htbp]
\center
\includegraphics[width=0.48\textwidth]{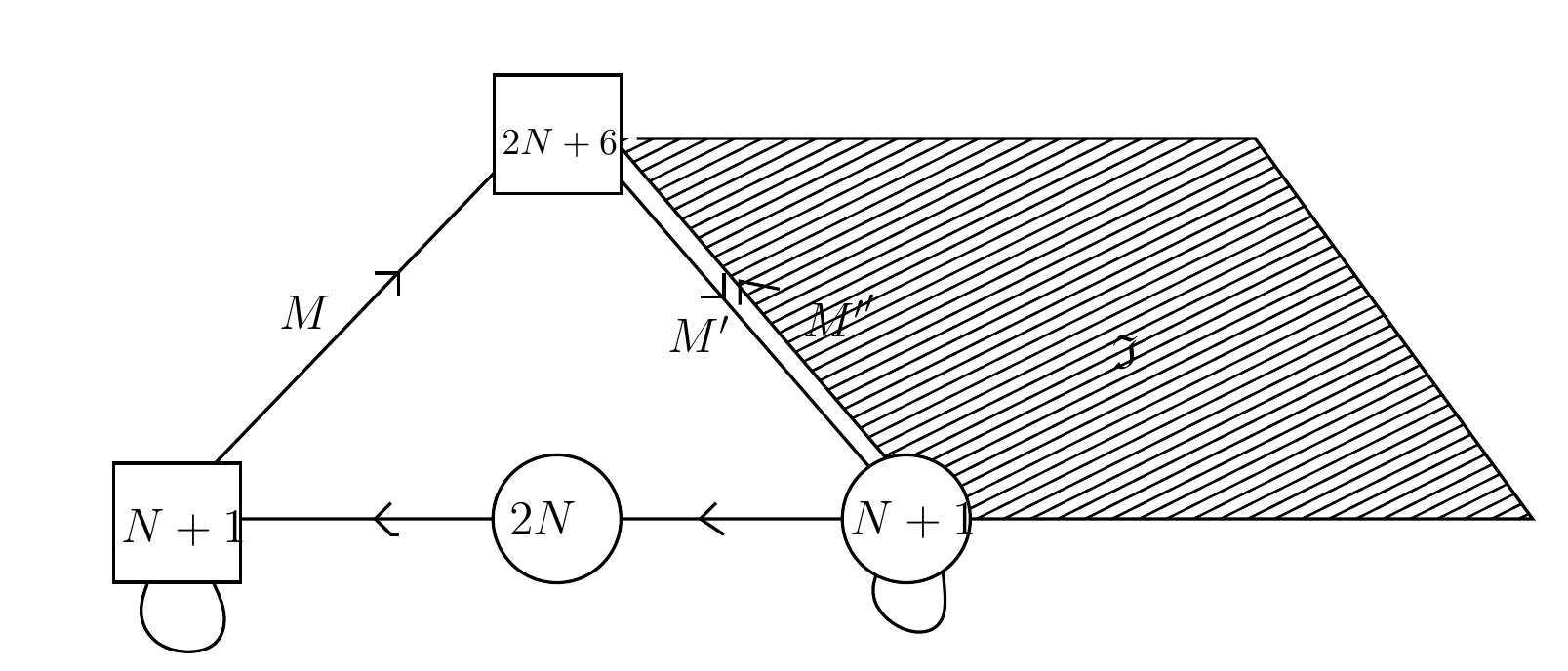} \includegraphics[width=0.42\textwidth]{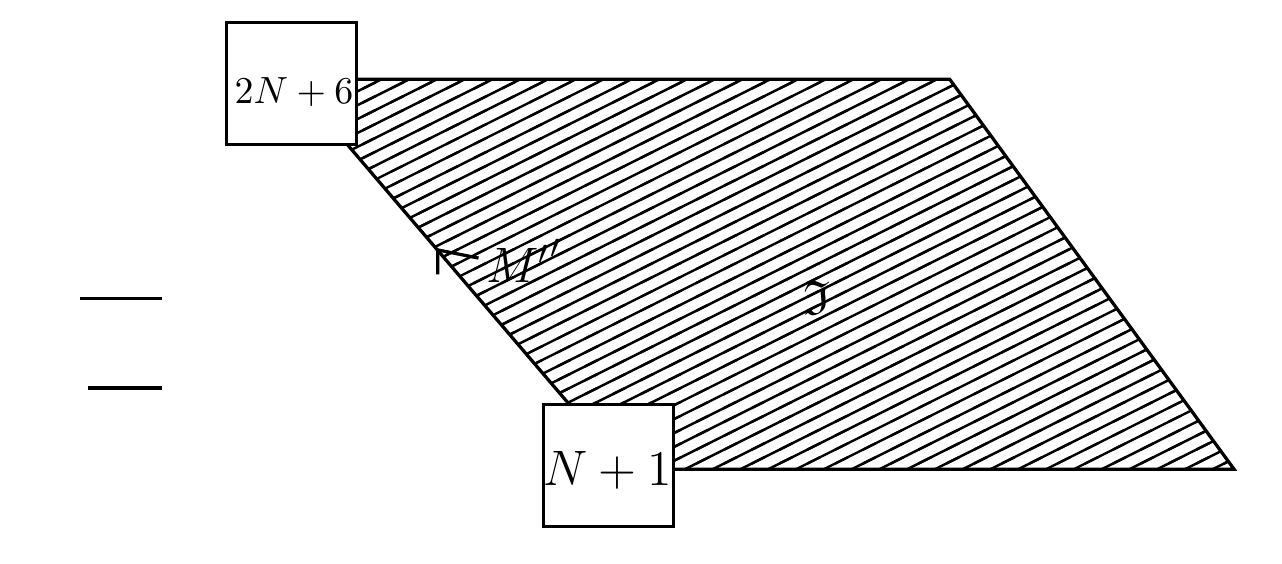} 
\caption{S gluing with $USp$ groups.}
\end{figure}

This is also consistent with flux domain walls in the $5d$ theory discussed in Section \ref{sec:5d}. A puncture is defined by boundary conditions for the $5d$ hypermultiplets. Different boundary conditions define different puncture types. S gluing corresponds to a special gluing : gluing two punctures with opposite boundary condition (or opposite sign). The gluing in this case is performed by gauging the diagonal flavor symmetry of two puncture symmetries which are either $SU(N+1)$ or $USp(2N)$. The resulting gauge theory connecting two punctures can be considered as the $5d$ conformal matter theory with $SU(N+1)$ or $USp(2N)$ gauge group on an interval with opposite boundary condition for hypermultiplets at two ends. Due to the opposite boundary conditions, the $5d$ hypermultiplets become massive which can effectively be described by the superpotential $M M'$ above for a hypermultiplet $\Phi=(M,M')$. This gluing for example occurs when we connect two domain walls with opposite flux like the second combination in (\ref{eq:two-domain-walls-flux}) for $SU(N+1)$ or like when $r=0$ for $USp(2N)$ in the second chamber of the flux wall configurations discussed in Section \ref{sec:5d}. When we put the $5d$ theory with this domain wall configuration on a tube, while preserving maximal puncture symmetry, the $5d$ theory reduces to the $4d$ tube model introduced above with either $SU(N+1)$ gauging or $USp(2N)$ gauging. Since two domain walls have opposite flux, the combination should give a trivial domain wall with zero flux. This implies the triviality of the $4d$ tube models with opposite flux in the $4d$ reduction.

Mathematically the equality of indices of ${\frak T}$ with the two tubes and ${\frak T}$ without them is due to the $(A,C)$ and $(C,A)$ inversion formulas of Spiridonov and Warnaar \cite{siwn}. In the case of $USp$ gauging the formula has a proof where in the case of $SU$ gauging it was conjectured.  Consistency of our picture is a physical motivation for such a conjecture to hold and it is related to the physical conjecture here of generating the vacuum expectation value by quantum effects.

\section{Affine quivers and duality taster}

As we mentioned in the discussion of the reduction of the $(D,D)$ conformal matter to five dimensions in addition to the $SU$ and $USp$ gauge theory description one can obtain a gauge theory with $SU(2)^N$ gauge theory. The quiver theory turns out to take the form of the affine Dynkin diagram of $D_{N+3}$\footnote{Naively the quiver appears to be linear and not shaped like a $D$ type Dynkin diagram. However, when the nodes in the center of the diagram are $SU(2)$ then the edge nodes need to be "$SU(1)$". From the study of brane webs and partition functions, it appears that these "$SU(1)$" factors should be interpreted as two fundamental hypermultiplets for the $SU(2)$ they are connected to. Thus, the $4$ flavors at the two ends play the role of the edge nodes. See section $2.2$ in \cite{HKLTY} and references there in.}. In fact this is the description which generalizes to $(ADE,ADE)$ conformal matter and we will discuss this in detail in a forthcoming publication \cite{affpuyt}.  All these different descriptions can be used to construct theories corresponding to torus compactifications with flux. In particular with same value of flux they should give equivalent conformal field theories in four dimensions. Let us here discuss a particular example of such an equivalence with the details of the derivation and generalizations postponed to \cite{affpuyt}.

\begin{figure}[htbp]
\center
\includegraphics[width=0.6\textwidth]{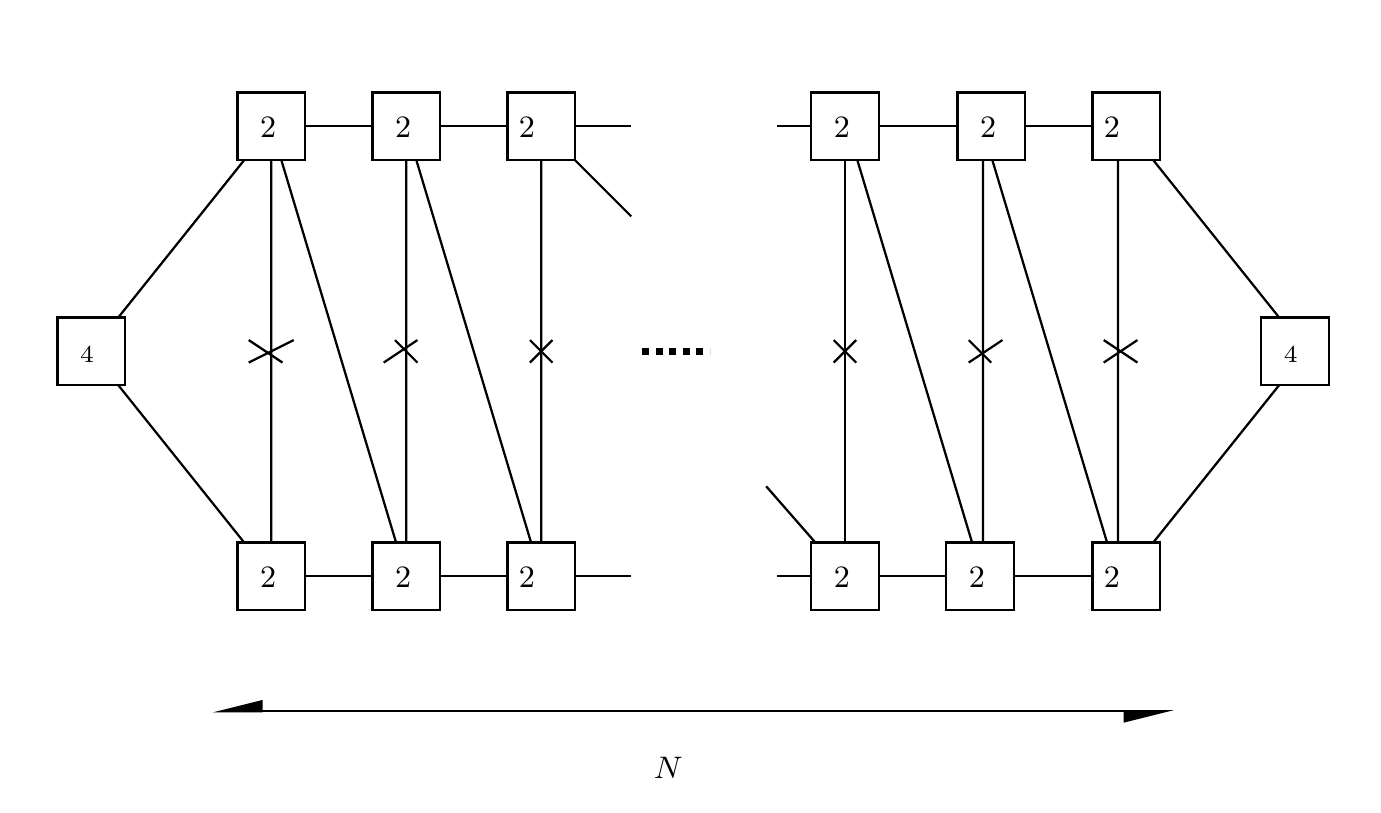} 
\caption{The tube with with $SU(2)^N$, affine,  maximal punctures.}
\label{tbttueu}
\end{figure}

The claim is that torus compactifications of $(D_{N+3},D_{N+3})$ minimal conformal matter with flux breaking the $SO(4N+12)$ symmetry to $SO(2N+10)\times U(1)\times SU(N+1)$ can be obtained by gluing together the Wess-Zumino model of Figure \ref{tbttueu}. This theory corresponds to flux of $1/(N+1)$ in the $U(1)$ direction. The gluing is performed by gauging the
 $SU(2)^N$ symmetry and adding bifundamental fields forming the affine Dynkin diagram of $D_{N+3}$, introducing bifundamental fields $\Phi_{i;i+1}$ for $i=1,...,N-1$ and two fields in fundamental of the last (first) $SU(2)$ and in fundamental of the first (second) $SU(4)$.
In plain words gluing is  identifying the edges of two glued Wess-Zumino models. 

\begin{figure} 
\center
\includegraphics[width=0.7\textwidth]{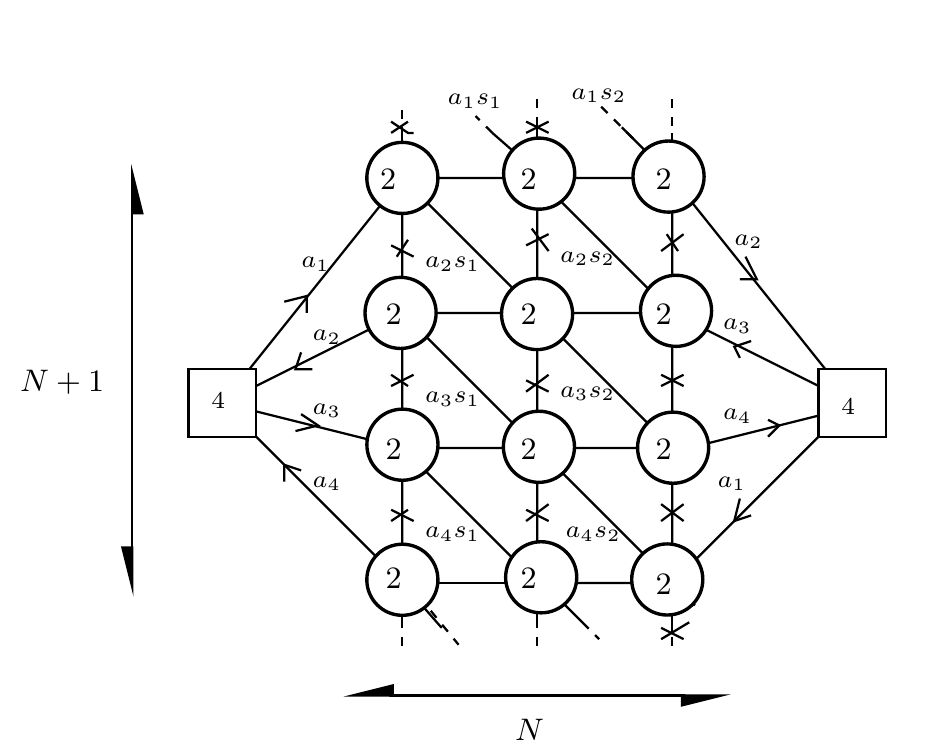} 
\caption{Torus with a unit of flux. The example here is for $N$ being three. All the fields are charged under $U(1)_t$ symmetry. The flipped fields have charge  one. The un-flipped fields get charge minus a half, and the flipping fields charge minus two.
 The two $SU(4)\sim  \, SO(6)$ symmetries together with $N-1$ $SO(2)_{s_i}$ symmetries build $SO(2N+10)$. We have $\prod_{l=1}^{N+1} a_l=1$ and this gives us $SU(N+1)$. The symmetry which has the flux is $U(1)_t$. The six dimensional R symmetry of the flip fields is two, of the flipped fields is zero, and of all other fields is one.}
 \label{dtryuh}
\end{figure}

Let us consider constructing a torus with flux one. To do so we need to glue $N+1$ copies of the Wess Zumino model, see Figure \ref{dtryuh}.  There are two different cases, when $N$ is even and when $N$ is odd. For $N$ even some of the rank of the symmetry is broken and we need to combine $2(N+1)$ copies to preserve all symmetry. In the case of $N$ odd we preserve all symmetry\footnote{This is related to the difference in quantization between even and odd $r=N+1$ discussed in section \ref{sec:6d}. Particularly for odd $r$, this flux is not consistent unless one also introduces a central flux in the $Z_2$ center of $SO(2N+10)$ that acts on the two spinors. This central flux exists in the tube for any $r$, but when connecting $r$ such tubes, is canceled out for $r$ even. There is also a central flux in the $Z_{N+1}$ center of $SU(N+1)$ which causes the breaking of some of that symmetry for torus compactification when the number of tubes is not an integer multiple of $N+1$.}.
Note that the symmetry one can see explicitly in the quiver is $SU(4)\times SU(4)\times U(1)^{2N}$. We claim that this symmetry enhances to $U(1)\times SO(2N+10)\times SU(N+1)$. 
The conformal R symmetry is the free one and the central charges are computed to be,

\be
a =\frac{(5N+1)(N+1)}6\, ,\qquad c= \frac{(11N+4)(N+1)}{12} \, .
\ee This agrees with the choice of $r=N+1$ for the $U(1)$ flux and with the charges of the model  of Figure \ref{Tubesa}. Note that the theory in that Figure and the one discussed here {\it look rather different but we claim that they should be dual}. That would be a non trivial consequence of different,  but dual in certain sense, five dimensional reductions of the $(D_{N+3}, D_{N+3})$  minimal conformal matter.
All the anomalies match between them and also match the computation from integrating anomaly polynomial from six dimensions, and also one can verify that indices match in examples.

 Let us quote the computation of the $N=3$ example.
 The supersymmetric operators form representations of $SU(4)\times SO(16)\times U(1)$. 
Some basic operators can be read off the quiver easily. For example, the flipping operators  form two copies of the rank two  antisymmetric of $SU(N+1)$ times a singlet of $SO(2N+10)$, 2$({\bf \frac{N(N+1)}{2}}  , \, {\bf 1})$. The quadratic gauge singlets in a single R charge one bifundamental field build $({\bf N+1} , {\bf 2N+10}) $ . The operators stretching between the ends of the quiver build two copies of the spinor ${\bf 2^{N+4}}$.
In the case of $N=3$ the index is using the conformal R symmetry and standard definitions of the index,

\be
&&1+\boxed{2({\bf 6}, \, {\bf 1}) t^{-2} (p q)^{\frac13}}+t^{-4} ( Sym^2 (2{\bf\;{\bf 6}) },\, {\bf 1}) (q p)^{\frac23}  +\boxed{t^{-1}({\bf 4},\, {\bf 16}) (q p  )^{\frac23}}+\nonumber\\
&& 2t^{-2}({\bf 6} ,\, {\bf 1})   (q+p)(q p)^{\frac13}+ t^{-3}2(({\bf \overline{20}},\,{\bf 16})+({\bf \overline { 4}}   ,\, {\bf 16})) q p+t^{-6} ( Sym^3 (2\;{\bf 6})   ,\, {\bf 1}) q p + \nonumber\\
&&t^{-1}({\bf 4},\, {\bf 16}) (q +  p) (q p  )^{\frac23}+t^{-4} (2({\bf 6}, \, {\bf 1}))\otimes (2({\bf 6}, \, {\bf 1}) )(q +  p)  (q p)^{\frac23}+2({\bf 6}, \, {\bf 1}) t^{-2}(q^2+p^2) (q p)^{\frac13} +  \nonumber\\ 
&&t^{-8} ( Sym^4 (2{\bf\;{\bf 6}) },\, {\bf 1}) (q p)^{\frac43}\boxed{-t ({\bf 4} ,\, {\bf 16}) (q p )^{\frac43}}+t^{-5}(Sym^2 (2{\bf\;{\bf 6}) }\otimes {\bf 4} ,\, {\bf 16}) ( q  p  )^{\frac43}+\\
&&+\boxed{2({\bf 1} ,\, {\bf 128}) t^{-2} (q p)^{\frac43}}+(Sym^2 ({\bf 4}  ,   \, {\bf 16}))t^{-2} (q p   )^{\frac43}+\boxed{3 t^{4} ({\bf 1} , \,  {\bf 1} ) (  p  q)^{\frac43}}\boxed{-t^{-2}2({\bf 4},{\bf 1}) (  q   p     )^{\frac43}}+ ... \nonumber
\ee The boxed terms are generators and the rest are products and derivatives.  Let us define the characters,

\be 
({\bf 4} ,\, {\bf 1}) =\sum_{j=1}^{N+1} a^2_j \, ,  \qquad     
 ( {\bf 1},  {\bf 2N+10} ) = \sum_{j\neq l}^4 \left\{(s^R_j s^R_l)^{\pm1} +(s^L_j s_l^L)^{\pm1}\right\} +\sum_{j=1}^{N-1} s^{\pm2}_j\,.
\ee Here $s^L (s^R)$ parametrize the left (right) $SU(4)$.

\begin{figure}[htbp]
\center
\includegraphics[width=0.6\textwidth]{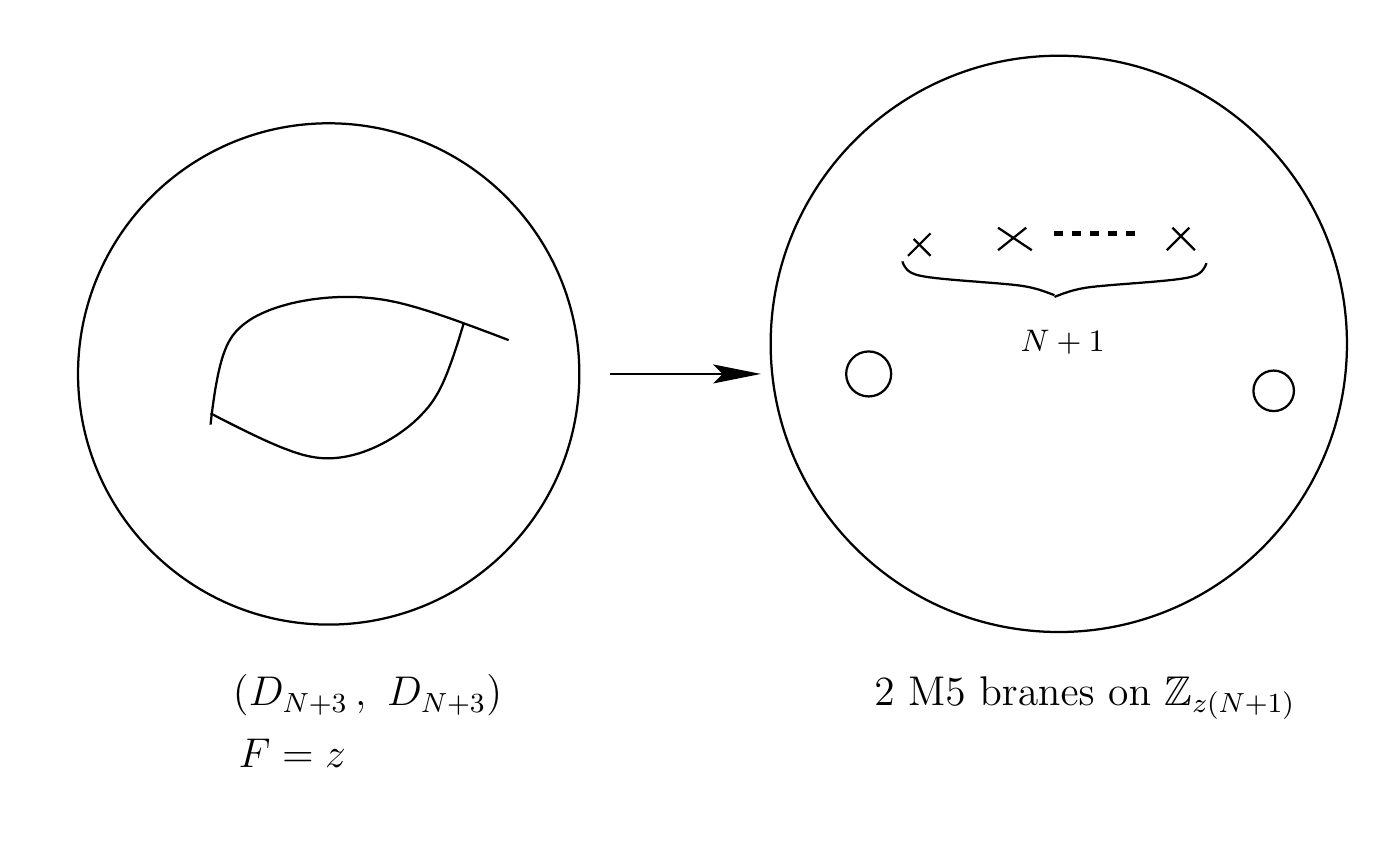} 
\caption{The relation between one M5 brane on D and two M5 branes on A types of singularity. }
\label{aandui}
\end{figure}

We also want to mention a connection of the torus models to models obtained from M5 branes probing A type singularity which can be deduced by observing our results and comparing them to \cite{Gaiotto:2015usa}. The statement will be that the models we find here, $(D_{N+3} \,  ,\, D_{N+3} )$ conformal matter on a torus  with flux $z$ for a particular $U(1)$ symmetry with $r=N+1$ is, up to flip fields, the same as two M5 branes probing ${\mathbb Z}_{(1+N)z}$ singularity on sphere with two maximal and $N+1$ minimal punctures. See Fig. \ref{aandui}.
It will be very interesting to understand whether there is an M/F theory explanation of such a relation.

\

\section{Branching Rules}

In this Appendix we summarize some branching rules that are useful in the study of compatifications of conformal matter with flux.
 These will be for the group $SO(4N+12)$ which is the global symmetry of the SCFT. In what follows we shall write the decomposition of the vector, adjoint and two spinors. The adjoint and spinors are useful as they appear in the $6d$ SCFT. The vector decomposition is mostly convenient to derive the decomposition of the characteristic classes. Throughout this section we shall use the following notations for representations: $V$ stands for the vector representation of an $SO$ group, $F, \overline{F}$ for the fundamental and anti-fundamental representations of an $SU$ group respectively, $\text{rank i AS}$ for the antisymmetric representation of rank $i$, $Ad$ for the adjoint representation, and $S, C$ for the two spinors of a $D_{2n}$ group.  

\subsection*{$SO(4N+12) \rightarrow U(1)\times SU(2N+6)$}

First we consider the $SO(4N+12) \rightarrow U(1)\times SU(2N+6)$ breaking. Here we have:

\be
V_{SO(4N+12)} \rightarrow F^{1}_{SU(2N+6)} + \overline{F}^{-1}_{SU(2N+6)},
\ee 

\bea
Ad_{SO(4N+12)} \rightarrow Ad^{0}_{SU(2N+6)} + (\text{rank 2 } AS)^{2}_{SU(2N+6)} + (\text{rank $2N+4$ } AS)^{-2}_{SU(2N+6)} + 1^{0}, \nonumber \\
\eea

\be
S_{SO(4N+12)} \rightarrow \sum^{N+3}_{i=-(N+3), \text{ $i$ even}} (\text{rank $N+3 + i$ } AS)^{i}_{SU(2N+6)},
\ee 

\be
C_{SO(4N+12)} \rightarrow \sum^{N+3}_{i=-(N+3), \text{ $i$ odd}} (\text{rank $N+3 + i$ } AS)^{i}_{SU(2N+6)} .
\ee

Note that the minimal charge here is normalized to $1$ and it is in the vector and $C$ spinor. In the adjoint and $S$ spinor the minimal charge is $2$. Since in $6d$ we have only an adjoint and one of the spinors, in one compactification, where the spinor we have in $6d$ decomposes as the $S$ spinor, the flux can be half-integer without relying on fractional fluxes.

In the special case of $N=2$, the complete decomposition is:

\be
\bold{20} \rightarrow \bold{10}^1 + \overline{\bold{10}}^{-1},
\ee 

\be
\bold{190} \rightarrow \bold{99}^0 + \bold{45}^2 + \overline{\bold{45}}^{-2} + \bold{1}^0 ,
\ee 

\be
\bold{512} \rightarrow \bold{10}^{-4} + \bold{120}^{-2} + \bold{252}^0 + \overline{\bold{120}}^{2} + \overline{\bold{10}}^{4} ,
\ee 

\be
\bold{512'} \rightarrow \bold{1}^{-5} + \bold{45}^{-3} + \bold{210}^{-1} + \overline{\bold{210}}^{1} + \overline{\bold{45}}^{3} + \bold{1}^{5} .
\ee 

\subsection*{$SO(4N+12) \rightarrow U(1)\times SU(r) \times SO(4N+12-2r)$}

For the generic case we have:

\be
V_{SO(4N+12)} \rightarrow (F_{SU(r)},1)^{1} + (\overline{F}_{SU(r)},1)^{-1} + (1,V_{SO(4N+12-2r)})^0, \label{VecDecomp}
\ee 

\bea 
Ad_{SO(4N+12)} \rightarrow & & (1,Ad_{SO(4N+12-2r)})^0 + (Ad_{SU(r)},1)^0 + (1,1)^0 + (\text{rank 2 } AS,1)^2 + (\text{rank $r-2$ } AS,1)^{-2} \nonumber \\  & & + (F_{SU(r)},V_{SO(4N+12-2r)})^1 + (\overline{F}_{SU(r)},V_{SO(4N+12-2r)})^{-1},
\eea 

\bea
&&S_{SO(4N+12)} \rightarrow (1,S_{SO(4N+12-2r)})^{-\frac{r}{2}} + (F_{SU(r)},C_{SO(4N+12-2r)})^{-\frac{r}{2}+1} +\nonumber\\
&&
\qquad\;\qquad  (\text{rank 2 } AS , S_{SO(4N+12-2r)})^{-\frac{r}{2}+2} + ..., \nonumber \\
\eea

\bea
&&C_{SO(4N+12)} \rightarrow (1,C_{SO(4N+12-2r)})^{-\frac{r}{2}} + (F_{SU(r)},S_{SO(4N+12-2r)})^{-\frac{r}{2}+1} +\nonumber\\
&&
\qquad\;\qquad
 (\text{rank 2 } AS , C_{SO(4N+12-2r)})^{-\frac{r}{2}+2} + ..., \nonumber \\
\eea  

Note that when $r$ is even the spinors of $SO(4N+12-2r)$ are real and we have two distinct self conjugate spinors, but for $r$ odd then $C_{SO(4N+12-2r)} = \overline{S}_{SO(4N+12-2r)}$, which is reflected in the decomposition.

It is interesting to note that here there are states with charge $1$ also in the adjoint. When $r$ is even this is indeed the minimal charge. However when $r$ is odd the minimal charge is $\frac{1}{2}$ which appears in the spinor. Therefore, if we adopt the uniform charge normalization  given by equation (\ref{VecDecomp}), then the flux is quantized so as to be half-integer, integer or even integer for the cases of $U(1)\times SU(2N+6)$ with spinor decomposing like $S$, $U(1)\times SU(2N+6)$ with spinor decomposing like $C$ or $r$ even and $r$ odd, respectively.  

Finally we shall write the decomposition for special cases that appear in this article. First for $N=2$, we consider the closely related cases of $r=8$ and $r=2$. 

\subsubsection*{$SO(20) \rightarrow U(1)\times SU(8) \times SO(4)$}

\be
\bold{20} \rightarrow (\bold{2},\bold{2},\bold{1})^0 + (\bold{1},\bold{1},\bold{8})^{1} + (\bold{1},\bold{1},\overline{\bold{8}})^{-1},
\ee 

\bea
\bold{190} \rightarrow (\bold{1},\bold{1},\bold{1})^0 + (\bold{3},\bold{1},\bold{1})^0 + (\bold{1},\bold{3},\bold{1})^0 + (\bold{1},\bold{1},\bold{63})^0 + (\bold{1},\bold{1},\bold{28})^2 + (\bold{1},\bold{1},\overline{\bold{28}})^{-2} + (\bold{2},\bold{2},\bold{8})^{1} + (\bold{2},\bold{2},\overline{\bold{8}})^{-1} , \nonumber \\
\eea 

\bea
& & \bold{512} \rightarrow (\bold{1},\bold{2},\bold{1})^{-4} + (\bold{2},\bold{1},\bold{8})^{-3} + (\bold{1},\bold{2},\bold{28})^{-2} + (\bold{2},\bold{1},\bold{56})^{-1} + (\bold{1},\bold{2},\bold{70})^{0} + (\bold{2},\bold{1},\overline{\bold{56}})^{1} + (\bold{1},\bold{2},\overline{\bold{28}})^{2} \nonumber \\ & & + (\bold{2},\bold{1},\overline{\bold{8}})^{3} + (\bold{1},\bold{2},\bold{1})^{4},
\eea 

\bea
& & \bold{512'} \rightarrow (\bold{2},\bold{1},\bold{1})^{-4} + (\bold{1},\bold{2},\bold{8})^{-3} + (\bold{2},\bold{1},\bold{28})^{-2} + (\bold{1},\bold{2},\bold{56})^{-1} + (\bold{2},\bold{1},\bold{70})^{0} + (\bold{1},\bold{2},\overline{\bold{56}})^{1} + (\bold{2},\bold{1},\overline{\bold{28}})^{2} \nonumber \\ & & + (\bold{1},\bold{2},\overline{\bold{8}})^{3} + (\bold{2},\bold{1},\bold{1})^{4} .
\eea 

Here we have ordered the global symmetry as $(SU(2),SU(2),SU(8))^{U(1)}$.

\subsubsection*{$SO(20) \rightarrow U(1)\times SU(2) \times SO(16)$}

\be
\bold{20} \rightarrow (\bold{2},\bold{1})^{1} + (\bold{2},\bold{1})^{-1} + (\bold{1},\bold{16})^{0},
\ee 

\be
\bold{190} \rightarrow (\bold{1},\bold{1})^0 + (\bold{1},\bold{1})^2 + (\bold{1},\bold{1})^{-2} + (\bold{3},\bold{1})^0 + (\bold{1},\bold{120})^0 + (\bold{2},\bold{16})^{1} + (\bold{2},\bold{16})^{-1} ,
\ee 

\be
\bold{512} \rightarrow (\bold{1},\bold{128})^{1} + (\bold{1},\bold{128})^{-1} + (\bold{2},\bold{128'})^{0},
\ee 

\be
\bold{512'} \rightarrow (\bold{1},\bold{128'})^{1} + (\bold{1},\bold{128'})^{-1} + (\bold{2},\bold{128})^{0}.
\ee 

Here we have ordered the global symmetry as $(SU(2),SO(16))^{U(1)}$.

Finally we consider the case of $N=3$, $r=4$.

\subsubsection*{$SO(24) \rightarrow U(1)\times SU(4) \times SO(16)$}

\be
\bold{24} \rightarrow (\bold{4},\bold{1})^{1} + (\overline{\bold{4}},\bold{1})^{-1} + (\bold{1},\bold{16})^{0},
\ee 

\be
\bold{276} \rightarrow (\bold{15},\bold{1})^0 + (\bold{1},\bold{1})^0 + (\bold{1},\bold{120})^0 + (\bold{6},\bold{1})^2 + (\bold{6},\bold{1})^{-2} + (\bold{4},\bold{16})^{1} + (\overline{\bold{4}},\bold{16})^{-1} ,
\ee 

\be
\bold{2048} \rightarrow (\bold{1},\bold{128'})^{-2} + (\overline{\bold{4}},\bold{128})^{-1} + (\bold{6},\bold{128'})^{0} + (\bold{4},\bold{128})^{1} + (\bold{1},\bold{128'})^{2},
\ee 

\be
\bold{2048'} \rightarrow (\bold{1},\bold{128})^{-2} + (\overline{\bold{4}},\bold{128'})^{-1} + (\bold{6},\bold{128})^{0} + (\bold{4},\bold{128'})^{1} + (\bold{1},\bold{128})^{2}.
\ee 

Here we have ordered the global symmetry as $(SU(4),SO(16))^{U(1)}$.

\bibliographystyle{ytphys}

\end{document}